\documentclass[fleqn,10pt, twocolumn,floatfix]{revtex4-2}
\usepackage[utf8]{inputenc}
\usepackage[T1]{fontenc}
\usepackage{amsmath}
\usepackage{amssymb}

\usepackage{mathtools}
\usepackage{xcolor}
\usepackage{subcaption}
\usepackage{verbatim}
\usepackage{float}
\usepackage{placeins}
\usepackage{bm}
\usepackage{subcaption}
\begin{document}
\title{Accelerating GW calculations through machine learned dielectric matrices}

\author{Mario G. Zauchner}
\affiliation{Department of Materials, Thomas Young Centre, Imperial College London, South Kensington Campus, London, SW7 2AZ, United Kingdom}
 
\author{Andrew Horsfield} 
\affiliation{Department of Materials, Thomas Young Centre, Imperial College London, South Kensington Campus, London, SW7 2AZ, United Kingdom}

\author{Johannes Lischner}
\email{j.lischner@imperial.ac.uk}
\affiliation{Department of Materials, Thomas Young Centre, Imperial College London, South Kensington Campus, London, SW7 2AZ, United Kingdom}

\begin{abstract}
The GW approach produces highly accurate quasiparticle energies, but its application to large systems is computationally challenging, which can  be largely attributed to the difficulty in computing the inverse dielectric matrix. To address this challenge, we develop a machine learning approach to efficiently predict density-density response functions (DDRF) in materials. For this, an atomic decomposition of the DDRF is introduced as well as the neighbourhood density-matrix descriptor both of which transform in the same way under rotations. The resulting DDRFs are then used to evaluate quasiparticle energies via the GW approach. This technique is called the ML-GW approach. To assess the accuracy of this method, we apply it to hydrogenated silicon clusters and find that it reliably reproduces HOMO-LUMO gaps and quasiparticle energy levels. The accuracy of the predictions deteriorates when the approach is applied to larger clusters than those included in the training set. These advances pave the way towards GW calculations of complex systems, such as disordered materials, liquids, interfaces and nanoparticles.
\end{abstract}

\flushbottom
\maketitle

\section{Introduction}
Density functional theory (DFT)\cite{Hohenberg1964, Kohn1965} has shown tremendous success in the calculation of electronic ground-state properties. However, it is well known that band gaps of solids and HOMO-LUMO gaps of molecules are often significantly underestimated when computed using Kohn-Sham (KS) eigenvalues \cite{Sham1983, Schultz2006}. In order to remedy this issue, the GW method \cite{Hedin1965,Strinati1982,Hybertsen1986} is often employed in which a self-energy correction to the DFT KS energies is computed. The resulting quasiparticle energies are in excellent agreement with experimental measurements for a wide range of materials. However, the large numerical effort required for GW calculations and its unfavorable scaling with system size restrict applications to relatively small systems~\cite{Onida2002, Rohlfing}. The most expensive step is the computation of the interacting density-density response function (DDRF) which is closely related to the inverse dielectric matrix. In particular, the non-interacting DDRF is typically computed by carrying out a slowly-converging summation over all unoccupied states~\cite{Adler, Wiser, Onida2002}. Afterwards, the non-interacting DDRF must be inverted to calculate the interacting DDRF.

These difficulties have led to the development of model DDRFs (or model dielectric matrices). For example, Hybertsen and Louie constructed a model dielectric matrix based on the assumption that the local screening response of the material is similar to that of a homogeneous medium with the same local density~\cite{Hybertsen}. A similar model was also proposed by Cappellini et al. \cite{Cappellini, BECHSTEDT1992765}. However, it has proven difficult to generalize these model dielectric functions to highly non-uniform systems, such as isolated molecules or nano-clusters whose screening properties are substantially different from uniform systems. To overcome this limitation, Rohlfing~\cite{Rohlfing} proposed to express the dielectric matrix as a sum of atomic contributions attributing a density response resulting from a Gaussian-shaped charge density to each atom. This model dielectric matrix contains a number of parameters which need to be determined, for example by comparison to calculated RPA dielectric functions. 

In recent years, machine learning (ML) techniques have been widely adopted to predict scalar properties of materials, such as the total energy. A key ingredient in ML approaches is the descriptor which parametrizes the atomic and chemical structure of the material. Many descriptors used in computational chemistry are explicitly constructed to be invariant under rotations and translations: for example, ACE \cite{Drautz2019}, SOAP \cite{Bartok2010}, the Coulomb matrix \cite{Hansen2012, Rupp2012}, bag-of-bonds \cite{hansen_machine_2015} or fingerprint-based descriptors have been shown to be reliable descriptors for the prediction of scalar quantities. When predicting tensors or functions, however, it is no longer sufficient to employ a rotationally invariant descriptor. To alleviate this problem, Grisafi et al.~\cite{Grisafi2018} developed a symmetry-adapted version of the SOAP kernel which is equivariant under rotations and was successfully used in the prediction of polarizability tensors and first hyperpolarizabilities \cite{Wilkins3401, Grisafi2018}, dipole moments \cite{Ceriotti2020} and electronic densities \cite{Grisafi2019}. Several other groups also explored ML approaches for the electronic density including Brockherde et al.\cite{brockherde_bypassing_2017}, Alred et al. \cite{alred_machine_2018} and Chandrasekaran and co-workers \cite{chandrasekaran_solving_2019}. Moreover, the construction of group-equivariant neural networks, such as Clebsch-Gordan networks \cite{kondor2018, pmlr-v80-kondor18a, Kondor2019}, tensor-field networks \cite{Thomas2018} and spherical convolutional neural networks (CNNs) \cite{cohen2018, cohen2016} have seen significant developments in recent years and the implementation of these methods has been significantly simplified by frameworks such as e3NN \cite{e3nn} developed by Geiger et al. \cite{e3nn_paper}, thus providing promising alternatives to the symmetry-adapted SOAP for the learning of functions.

To the best of our knowledge, however, there has been no attempt to develop ML models for the prediction of non-local response functions, such as the DDRF. Predicting such quantities is a formidable challenge: for example, the DDRF of a small silicon cluster can be tens of gigabytes in size when represented in a plane-wave basis even when a modest plane-wave cutoff is used. To address this problem, we introduce a decomposition of the DDRF into atomic contributions which can be predicted using ML techniques. To ensure that the ML model appropriately incorporates the transformation properties of the DDRF, we also develop a new descriptor called neighbourhood density-matrix (NDM) which transforms in the same way as the DDRF under rotations and is used in conjunction with a dense neural network to predict the atomic contributions to the DDRF. We then use the ML DDRFs to carry out GW calculations of hydrogenated silicon clusters. This approach which we refer to as the ML-GW method produces accurate GW quasiparticle energies at a significantly reduced computational cost compared to standard implementations.

\section{Results}
\subsection{Theoretical results}

The GW method yields accurate quasiparticle energies by applying a self-energy correction to the mean-field KS energy levels. 
The GW self-energy $\Sigma(\mathbf{r,r'}, \omega)$ is calculated from the one-electron Green's function $G(\mathbf{r,r'}, \omega)$ and the screened Coulomb interaction $W(\mathbf{r,r'}, \omega)$ according to~
\cite{Onida2002, Hybertsen1986, Louie1985}

\begin{equation}
    \Sigma(\mathbf{r,r'}, \omega) =  \frac{i}{2\pi} \int e^{-i\delta \omega'}G(\mathbf{r,r'}, \omega + \omega')W(\mathbf{r,r'},\omega')d\omega'
\end{equation}
with $\delta$ denoting a positive infinitesimal. The screened Coulomb interaction is in turn computed from the bare Coulomb interaction $v(\mathbf{r,r'})$ and the inverse dielectric matrix $\epsilon^{-1}(\mathbf{r,r'}, \omega)$ via
\begin{equation}
    W(\mathbf{r,r'}, \omega) = \int  \epsilon^{-1} (\mathbf{r},\mathbf{r}_2, \omega)v(\mathbf{r}_2,\mathbf{r'}) d\mathbf{r}_2,
\end{equation}
which demonstrates that the dielectric matrix constitutes a key ingredient in GW calculations. It can be obtained from the interacting DDRF $\chi(\mathbf{r},\mathbf{r}',\omega)$ according to
\begin{equation}
    \epsilon^{-1} (\mathbf{r,r'}, \omega) =\delta(\mathbf{r,r'}) + \int v(\mathbf{r},\mathbf{r}_2)\chi(\mathbf{r}_2,\mathbf{r'}, \omega)d\mathbf{r}_2.
\end{equation}
In the remainder of this paper, we will assume that the frequency dependence of the dielectric matrix can be approximated by the generalized plamon-pole approximation (GPP) \cite{Hybertsen1986,Lischner2014, Sharifzadeh2012}. As a consequence, only the static DDRF $\chi(\mathbf{r},\mathbf{r}') \equiv \chi(\mathbf{r},\mathbf{r}',\omega=0)$ needs to be determined. 

Within the random-phase approximation (RPA), the interacting static DDRF is given by 
\begin{align}
        \chi(\mathbf{r},\mathbf{r}')=&\chi_0(\mathbf{r},\mathbf{r}') + \\ &\int d\mathbf{r}_1 d\mathbf{r}_2 \chi_0(\mathbf{r},\mathbf{r}_1) v(\mathbf{r}_1,\mathbf{r}_2) \chi(\mathbf{r}_2,\mathbf{r}')
        \label{dyson}
\end{align}
with $\chi_0(\mathbf{r},\mathbf{r}')$ denoting the static non-interacting DDRF which is typically computed as a sum over empty and occupied states~\cite{Adler, Wiser} according to 
\begin{align}
        \chi_0(\mathbf{r},\mathbf{r}')=&\sum_{ij} \frac{f_i(1-f_j)}{\epsilon_i-\epsilon_j} \times \\ &\left[ \phi_i^*(\mathbf{r})\phi_j(\mathbf{r})  \phi^*_j(\mathbf{r}')\phi_i(\mathbf{r}') + \text{c.c.} \right].
        \label{adlerwiser}
\end{align}
Here, $\epsilon_i$, $f_i$ and $\phi_i(\mathbf{r})$ denote the orbital energy, occupancy and wavefunctions of the KS state $i$. 

Equations~\eqref{dyson} and \eqref{adlerwiser} highlight the two main challenges in computing the DDRF: (1) the calculation of the non-interacting DDRF requires a summation of all empty states which is slowly converging and (2) the calculation of the interacting DDRF requires a matrix inversion which scales unfavorably with system size.

\subsubsection{Atomic decomposition of the density-density response function}
In order to bypass the expensive computation of the DDRF and pave the way towards a machine learning approach, we propose to express 
$\chi(\mathbf{r,r'})$ as a sum of atomic contributions $\chi_i(\mathbf{r,r'})$ according to
\begin{equation}
    \chi(\mathbf{r,r'}) = \sum_{i=1}^N \chi_i(\mathbf{r,r'}),
\end{equation}
where $i$ labels atoms and $N$ is the total number of atoms. 

How this partitioning is achieved is not immediately obvious. However, the atomic contributions to the DDRF should have the following properties: (1) the atomic contributions should be localized in the vicinity of the corresponding atom, (2) they should retain the global symmetry of $\chi$, i.e. $ \chi(\mathbf{r,r'}) = \chi(\mathbf{r',r})$, and (3) they should integrate to zero, i.e. $\int \chi(\mathbf{r},\mathbf{r}') d\mathbf{r}=\int \chi(\mathbf{r},\mathbf{r}') d\mathbf{r'}=0$, to ensure that the change in the charge density induced by a perturbing potential is overall charge neutral~\cite{Onida2002}.

We start by expressing the DDRF in a localized basis set of real orbitals $\{\phi_{\alpha_a}^{a}(\mathbf{r}) \}$, where $a$ labels the atom on which the basis function is centered and $\alpha_a$ indexes the orbital on site $a$. We should note that exploiting locality
in GW calculations through local orbital representations
has been done before, even as early as the first practical
applications of the GW method by Strinati et al.~\cite{Strinati1982}. In this basis the DDRF is given by \begin{equation}
    \chi(\mathbf{r},\mathbf{r}') = \sum_{a, \alpha_a}   \sum_{b, \alpha_b} \chi_{\alpha_a \alpha_b}^{ab}
  \phi_{\alpha_a}^{a}(\mathbf{r})  \phi_{\alpha_b}^{b}(\mathbf{r}'),
  \label{intermediate}
\end{equation}
where $\chi_{\alpha_a \alpha_b}^{ab}$ is a symmetric matrix. This expression suggests the following decomposition of the DDRF into atomic contributions 
\begin{multline}
 \chi_i(\mathbf{r,r'}) = 
    \frac{1}{2} \sum_{\alpha_i} \sum_{b, \alpha_b} \bigg(\chi^{ib}_{\alpha_i\alpha_b}     \phi_{\alpha_i}^{i} (\mathbf{r})     
        \phi_{\alpha_b}^{b}( \mathbf{r'})  \\  +   \chi^{bi}_{\alpha_b\alpha_i}     \phi_{\alpha_b}^{b}(\mathbf{r})     \phi_{\alpha_i}^{i} (\mathbf{r'})\bigg ).
        \label{atomiccontribution}
\end{multline}
We refer to the representation of the DDRF in the basis $\{ \phi^a_{\alpha_a}(\mathbf{r})\}$ as 2-center DDRF (2C-DDRF) because it contains pairs of basis functions which are centered on different atoms.

Using the symmetry of $\chi^{iw}_{\alpha_i\alpha_w}$ and the fact that the basis functions are real, it can be easily verified that $  \chi_i(\mathbf{r,r'})=\chi_i(\mathbf{r',r}) $. We can also ensure that $ \int \chi_i(\mathbf{r},\mathbf{r}')d\mathbf{r}=0$ by removing all s-orbitals from the basis: see computational methods section for details.
The locality of $ \chi_i(\mathbf{r,r'})$ is directly inherited from the corresponding properties of the full DDRF. In particular, we have found that the expansion coefficients $\chi^{iw}_{\alpha_i\alpha_w}$ decay rapidly as the distance between atom $i$ and atom $w$ increases~\cite{MUSSARD201544}. 

We stress that this atomic representation of the DDRF is exact, i.e. $\sum_i \chi_i(\mathbf{r},\mathbf{r}')$ reproduces the full interacting DDRF when the local basis sets is complete. However, the atomic contributions to the DDRF contain contributions from pairs of basis functions which are centered on different atoms, see Eq.~\eqref{atomiccontribution}. These contributions are difficult to learn using atom-centered descriptors. 

To make progress, we exploit the localization of $\chi_i(\mathbf{r},\mathbf{r}')$ and expand it in terms of a set of basis functions $\psi^i_{nlm}(\mathbf{r})=Y_{lm}(\hat{\mathbf{r}})R_n(\vert \mathbf{r}\vert)$ (with $Y_{lm}$ denoting the spherical harmonics and $R_n$ a set of radial functions) which are all centered on atom $i$ according to
\begin{multline}
    \chi_i(\mathbf{r,r'}) =  \sum_{nlm}\sum_{n'l'm'} \chi^{(i)}_{nlmn'l'm'}\\ Y_{lm}(\hat{\mathbf{r}})Y_{l'm'}^{*}(\hat{\mathbf{r}}')R_n(\vert \mathbf{r} \vert)R_{n'}^{*}(\vert \mathbf{r'} \vert) 
    \label{chi_atomic}
\end{multline}
with $\chi^{(i)}_{nlmn'l'm'}$ denoting the expansion coefficients given by
\begin{multline}
    \chi^{(i)}_{nlmn'l'm'} = \int \int d\mathbf{r} d\mathbf{r'} \chi_i(\mathbf{r,r'}) R^*_n(\vert \mathbf{r} \vert)R_{n'}(\vert \mathbf{r'}\vert ) \\Y^*_{lm}(\mathbf{\hat{r}})Y_{l'm'}(\mathbf{\hat{r}'}).
    \label{atomicdec}
 \end{multline}
These coefficients can be learned using a neural network based on atom-centered descriptors. We refer to the representation of the DDRF in the basis $\{ \psi^i_{nlm}(\mathbf{r})\}$ as 1-center DDRF (1C-DDRF) because it only contains pairs of basis functions centered on the same atom.

\subsubsection{Neighbourhood density-matrix descriptor}
As discussed in the introduction, it is not appropriate to use a scalar descriptor (such as the standard SOAP descriptor~\cite{Bartok2013}) that is invariant under rotations to develop a ML model for the DDRF: the behaviour of the atomic DDRFs under rotations is determined by their analytical form: see Eq.~\eqref{chi_atomic}. In particular, we show in the Appendix that the coefficients of the atomic DDRF transform according to 
\begin{equation}
\tilde{\chi}^{(i)}_{nlm_1n'l'm_2} =      \sum_{m, m'} D^{l}_{m_1m}(\hat{R})D^{l'*}_{m_2m'}(\hat{R}) \chi^{(i)}_{nlmn'l'm'},
\end{equation}
where $\tilde{\chi}^{(i)}_{nlmn'l'm'}$ denote the coefficients of the tranformed DDRF, $\hat{R}$ is a rotation and $D^l_{mm'}(\hat{R})$ is a Wigner D-matrix~\cite{rose2013elementary}. 

Next, we construct the NDM descriptor, which transforms under rotations in the same way as the atomic DDRF. The starting point for such a descriptor is a non-local extension of the smooth neighbourhood density of atom $i$ of species $\eta$ employed in the SOAP descriptor \cite{Bartok2010}, defined as
\begin{equation}
\rho^{ \eta}_i(\mathbf{r,r'}) =\sum_{k \in \eta} \sum_{l \in \eta}  e^{-\alpha (\mathbf{r}-\mathbf{r}_k)^2}e^{-\alpha (\mathbf{r}'-\mathbf{r}_l)^2},
    \label{desc}
\end{equation}
where $k$ and $l$ run over atoms in the neighbourhood of atom $i$ within a cut-off radius $R_{cut}$ and $\alpha$ is a hyperparameter which describes the size of an atom. The NDM is then expanded in a basis of spherical harmonics and radial basis functions $R_n(|\mathbf{r}|)$ according to 
\begin{multline}
    \rho^{ \eta}_i(\mathbf{r,r'})  = \sum_{nlm}\sum_{n'l'm'} \rho^{(i, \eta)}_{nlmn'l'm'} \\
    Y_{lm}(\hat{\mathbf{r}})Y_{l'm'}^{*}(\hat{\mathbf{r}}')R_n(|\mathbf{r}|)R_{n'}^{*}(|\mathbf{r'}|),
\end{multline}
with $\rho^{(i, \eta)}_{nlmn'l'm'}$ being expansion coefficients. The above equation shows that the NDM transforms in the same way as the atomic DDRF: see Appendix for additional details. Therefore, we use the expansion coefficients as a descriptor for learning the DDRF. 

We note that the NDM can be written as the product of two neighbourhood densities $\rho^\eta_i(\mathbf{r})=\sum_{k \in \eta}  \exp\{-\alpha {(\mathbf{r}-\mathbf{r}_k)^2}\}$ according to
\begin{equation}    \rho^{\eta}_i(\mathbf{r,r'}) = \rho^\eta_i(\mathbf{r})\rho^\eta_i(\mathbf{r}').
\end{equation}
Similar to the NDM, $\rho^\eta_i(\mathbf{r})$ can be expanded in a basis of spherical harmonics and radial basis functions $R_n(|\mathbf{r}|)$ with coefficients $\rho^{(i,\eta)}_{nlm}$. It follows that 
\begin{equation}
    \rho^{(i,\eta)}_{nlmn'l'm'}=\rho^{(i,\eta)}_{nlm}\rho^{(i,\eta)}_{n'l'm'},
\end{equation}
which demonstrates that the coefficients of the neighbourhood density contain the same information as the coefficients of the neighbourhood density matrix. Indeed, we have found in our calculations that both types of coefficients perform equally when used as descriptors to predict the atomic DDRFs. 
We further note that the coefficients of the 3-body version of the SOAP descriptor $d^{(\eta)}_{nn'l}$ can be obtained from the NDM using
\begin{equation}
    d^{(\eta)}_{nn'l} = \sum_{l'mm'}\sqrt{\frac{8\pi^2}{2l+1}}\rho^{(i,\eta)}_{nlm}\rho^{(i,\eta)}_{n'l'm'} \delta_{ll'}\delta_{mm'},
\end{equation}
in the case where there is no coupling between different atomic species $\eta$.
\subsection{Machine learning}
We apply our ML approach for predicting DDRFs to hydrogenated silicon clusters and then use the DDRFs to calculate GW quasiparticle energies for these systems. We refer to this technique as the ML-GW approach. The atomic positions of the clusters were constructed as described in the methods section and then relaxed using DFT. 

\begin{figure}[h]
    \centering

    \includegraphics[width=0.5\textwidth]{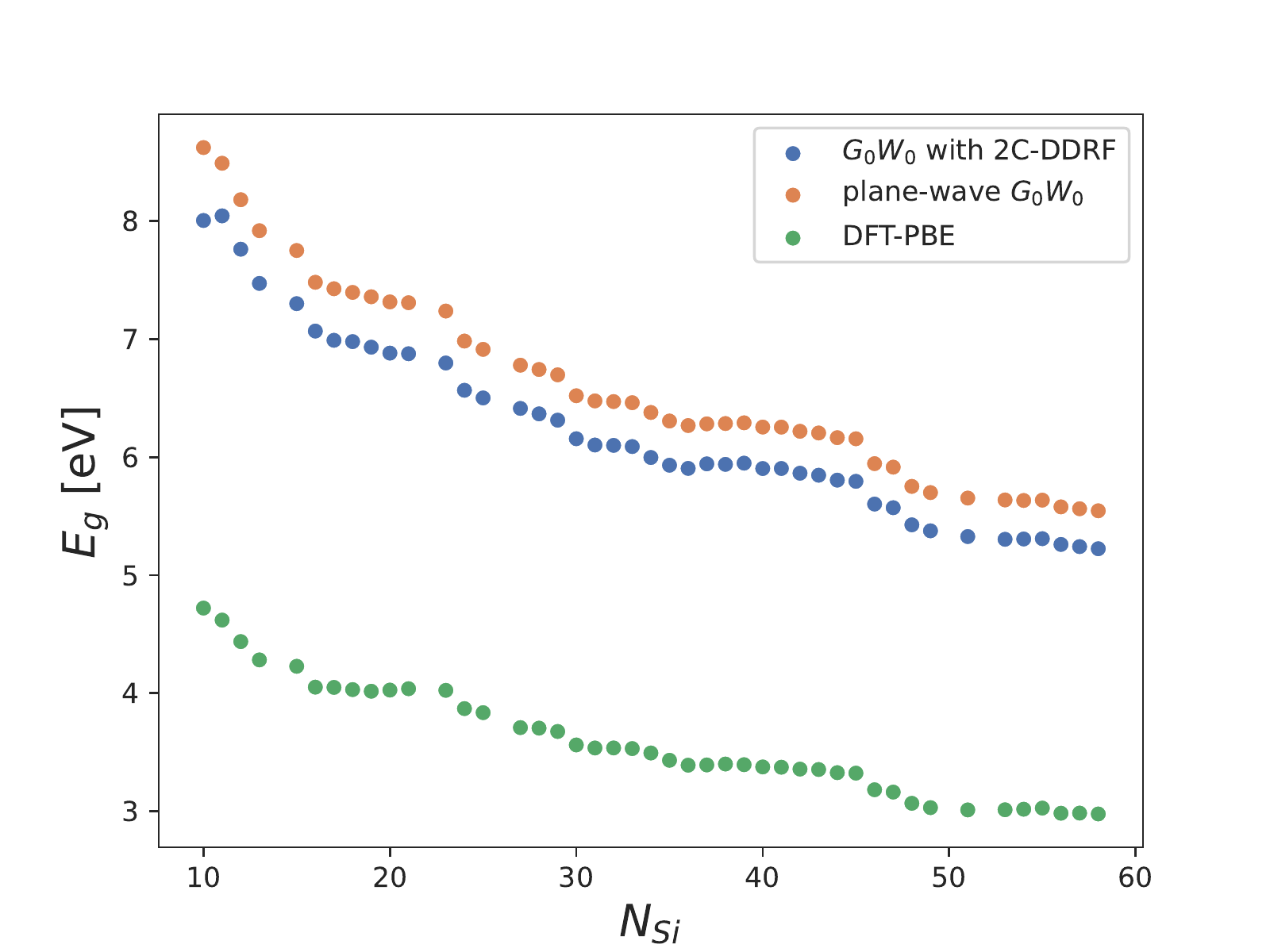}

     \caption{HOMO-LUMO gaps of hydrogenated silicon clusters from DFT-PBE Kohn-Sham eigenvalues, plane-wave G$_0$W$_0$ and G$_0$W$_0$ calculations using the 2C-DDRF, see section "Atomic decomposition of the density-density response function".}
    \label{crosssite}
\end{figure}

To establish the accuracy of this approach, we first investigate the error in the GW quasiparticle energies resulting from the expansion of the DDRF in terms of the intermediate local basis $\{ \phi^a_{\alpha_a}(\mathbf{r}) \}$: see Eq.~\eqref{intermediate}. Fig.~\ref{crosssite} compares the HOMO-LUMO gaps obtained from mean-field DFT-PBE calculations, a standard plane-wave G$_0$W$_0$ calculation using a generalized plasmon-pole approximation~\cite{Hybertsen1986,Sharifzadeh2012} and a G$_0$W$_0$ calculation using the 2C-DDRF, where the DDRF is expanded in terms of a modified version of the admm-2 basis set~\cite{Kumar2018}: see methods section. The DFT-PBE results show that the HOMO-LUMO gap decreases with increasing cluster size from $E_g \approx 4.8$~eV for the smallest cluster containing 10 Si atoms to $E_g \approx 3$~eV for the biggest cluster with almost 60 Si atoms. This decrease is a consequence of quantum confinement effects which are less pronounced for bigger clusters. The plane-wave GW HOMO-LUMO gaps show a similar trend as function of cluster size, but the gaps are larger than the DFT-PBE gaps by several electron volts. Interestingly, the GW corrections are larger for smaller clusters than for larger clusters. As a consequence, the reduction in the GW HOMO-LUMO gaps as a function of cluster size is larger compared to the DFT-PBE result: in particular, the gap is as large as 8.6 eV for the smallest clusters and shrinks to 5.5 eV for the largest clusters corresponding to a decrease of 3.1 eV (compared to a decrease of 1.8 eV in the DFT-PBE HOMO-LUMO gap energies). Similar results were obtained by Chelikowsky et al.~\cite{Chelikowsky} who also carried out GW calculations hydrogenated Si clusters. In particular, they found that the HOMO-LUMO gap shrinks from $\sim 9$~eV for a 10 Si atom cluster to $\sim 6.5$~eV for a 47 Si atom cluster. The GW results obtained with the 2C-DDRF are qualitatively similar to the plane-wave GW results. However, the HOMO-LUMO gaps that are obtained with this approach are consistently $\sim 0.4$~eV smaller than the plane-wave results. This is a consequence of the incompleteness of the local basis set

\begin{figure}[h]

    \centering
 \begin{subfigure}[t]{0.47\textwidth}   
 \caption{}   
 \includegraphics[width=\textwidth]{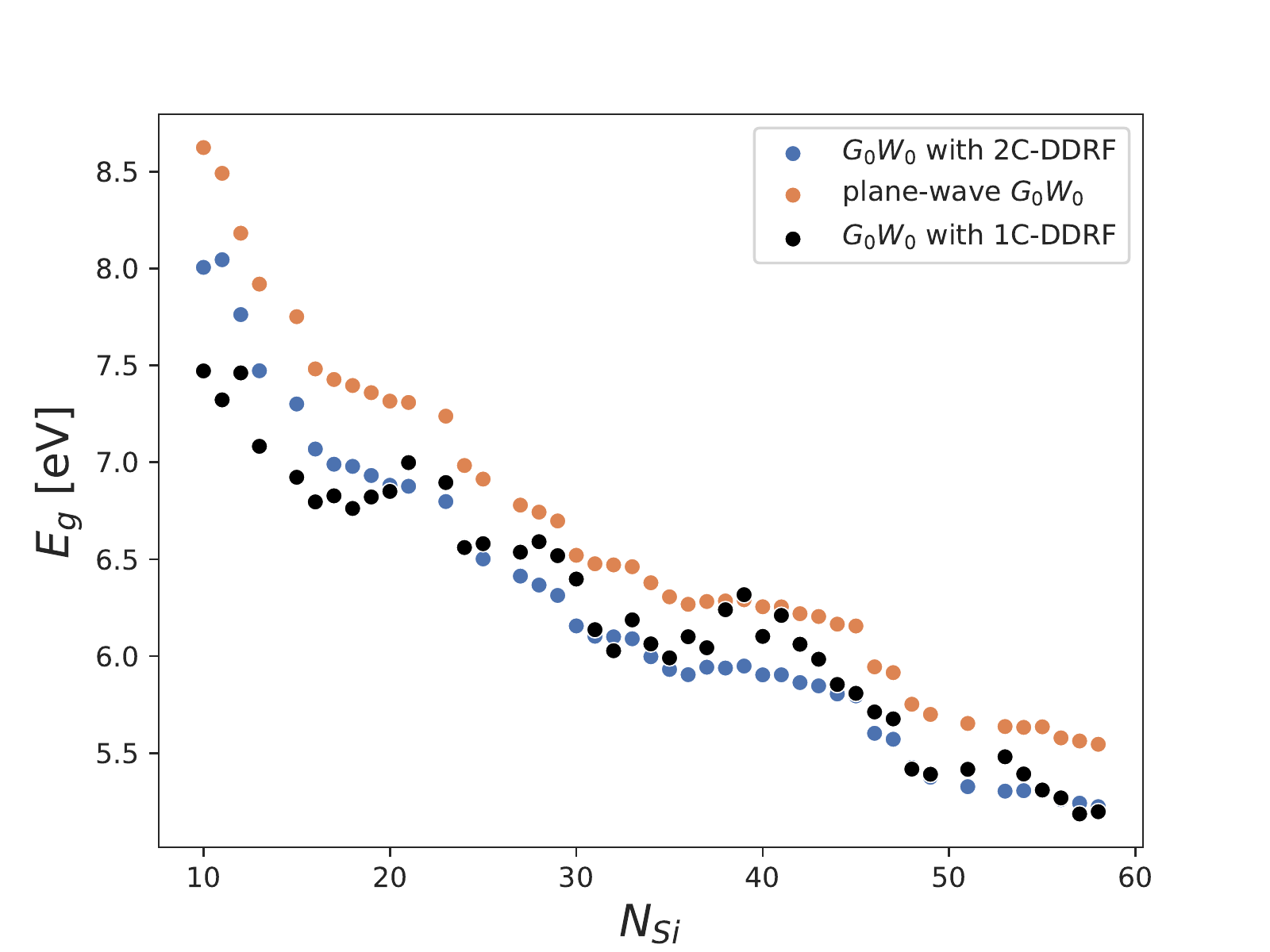}
        \end{subfigure}    \centering
         
\begin{subfigure}[t]{0.47\textwidth}   \caption{}
    \includegraphics[width=\textwidth]{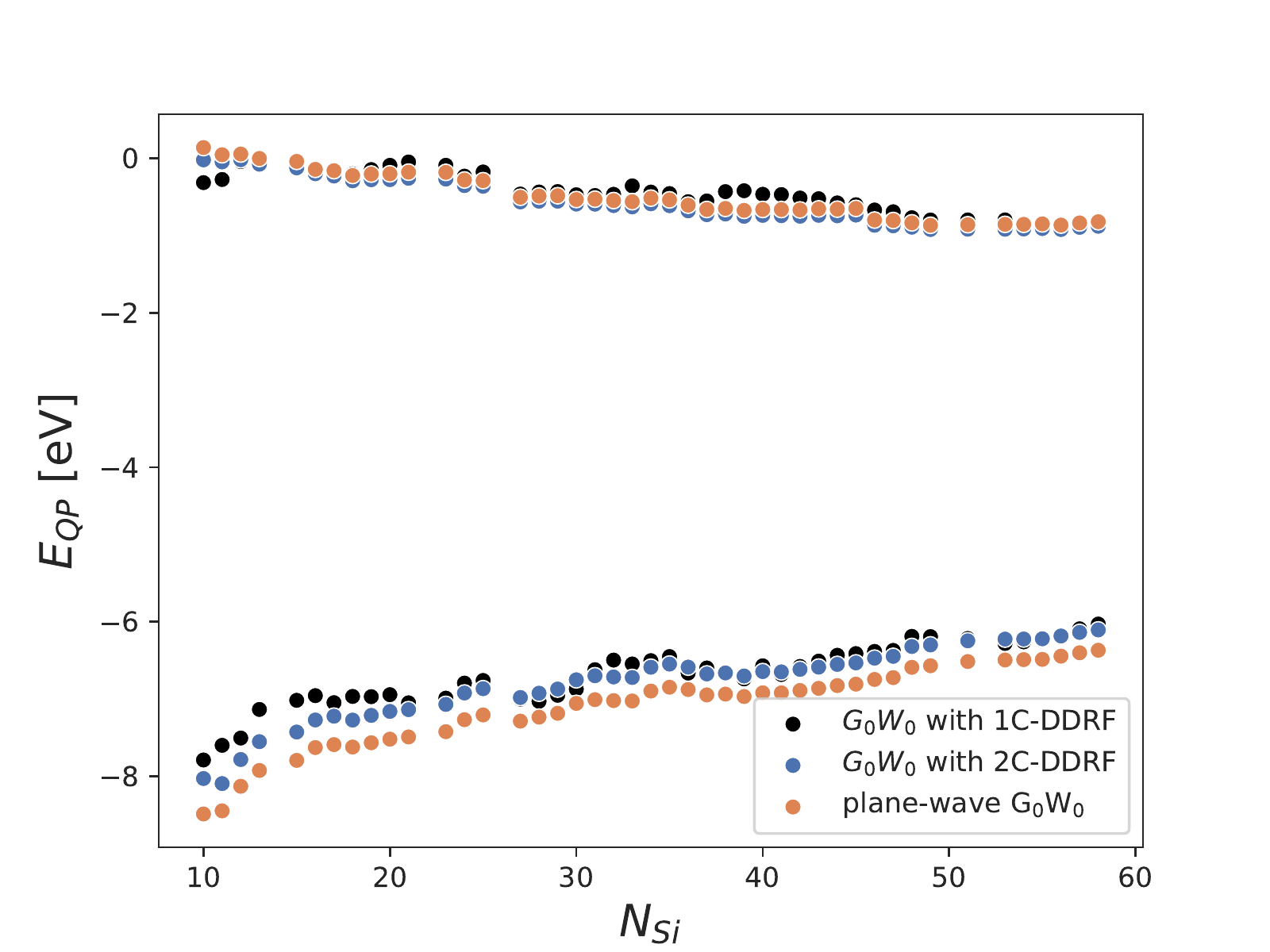}
        \end{subfigure}
     \caption{(a) HOMO-LUMO gaps of hydrogenated silicon clusters from plane-wave G$_0$W$_0$ and G$_0$W$_0$ calculations using the 2C-DDRF and G$_0$W$_0$ calculations using the 1C-DDRF, see section "Atomic decomposition of the density-density response function". (b) HOMO and LUMO energies of hydrogenated Si clusters.}
    \label{atomic}
\end{figure}

Next, we determine the 1C-DDRF. For the basis set we use solid harmonic Gaussians with optimized decay coefficients: see methods section. Fig.~\ref{atomic} (a) compares the HOMO-LUMO gaps from G$_0$W$_0$ calculations with the 1C-DDRF to those obtained with the 2C-DDRF and also to plane-wave G$_0$W$_0$ results. For small clusters, the HOMO-LUMO gaps obtained with the 1C-DDRF are smaller than those obtained with the 2C-DDRF, while the opposite behaviour is observed for larger cluster. The largest difference between the two methods is obtained for clusters containing $\sim 40$ Si atoms. The root-mean-square error (RMSE) of the 1C-basis results relative to the 2C-basis results is 0.22~eV and the RMSE relative to the plane-wave results is 0.45~eV for all clusters. Fig.~\ref{atomic} (b) shows the HOMO and LUMO quasiparticle energies. It can be seen that better agreement with the plane-wave result is obtained for the LUMO than for the HOMO.

\begin{figure}[h]    \centering
\begin{subfigure}[t]{0.47\textwidth}   \caption{}
    \includegraphics[width=\textwidth]{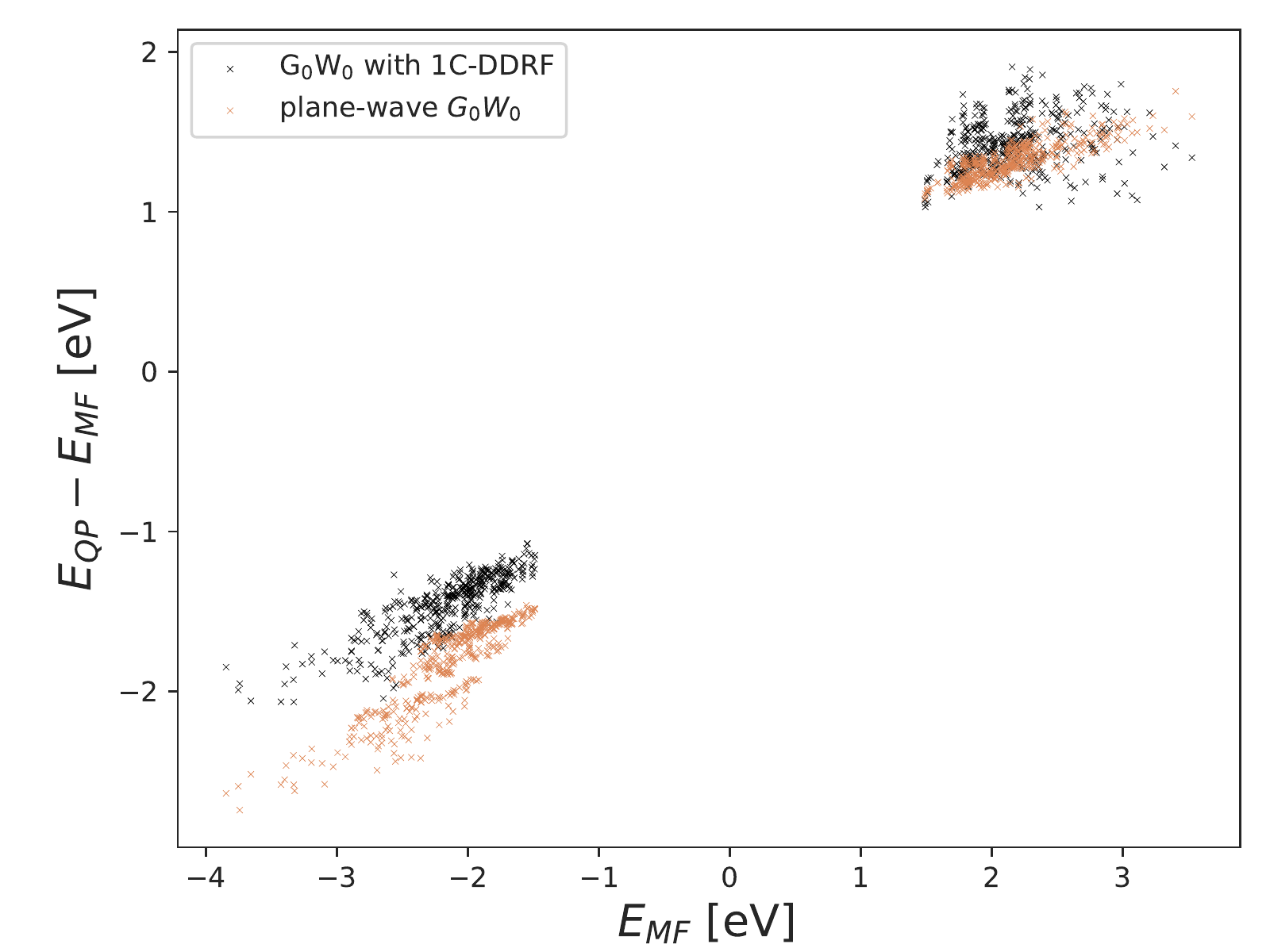}
        \end{subfigure}

\begin{subfigure}[t]{0.47\textwidth}
\caption{}
      \centering    \includegraphics[width=\textwidth]{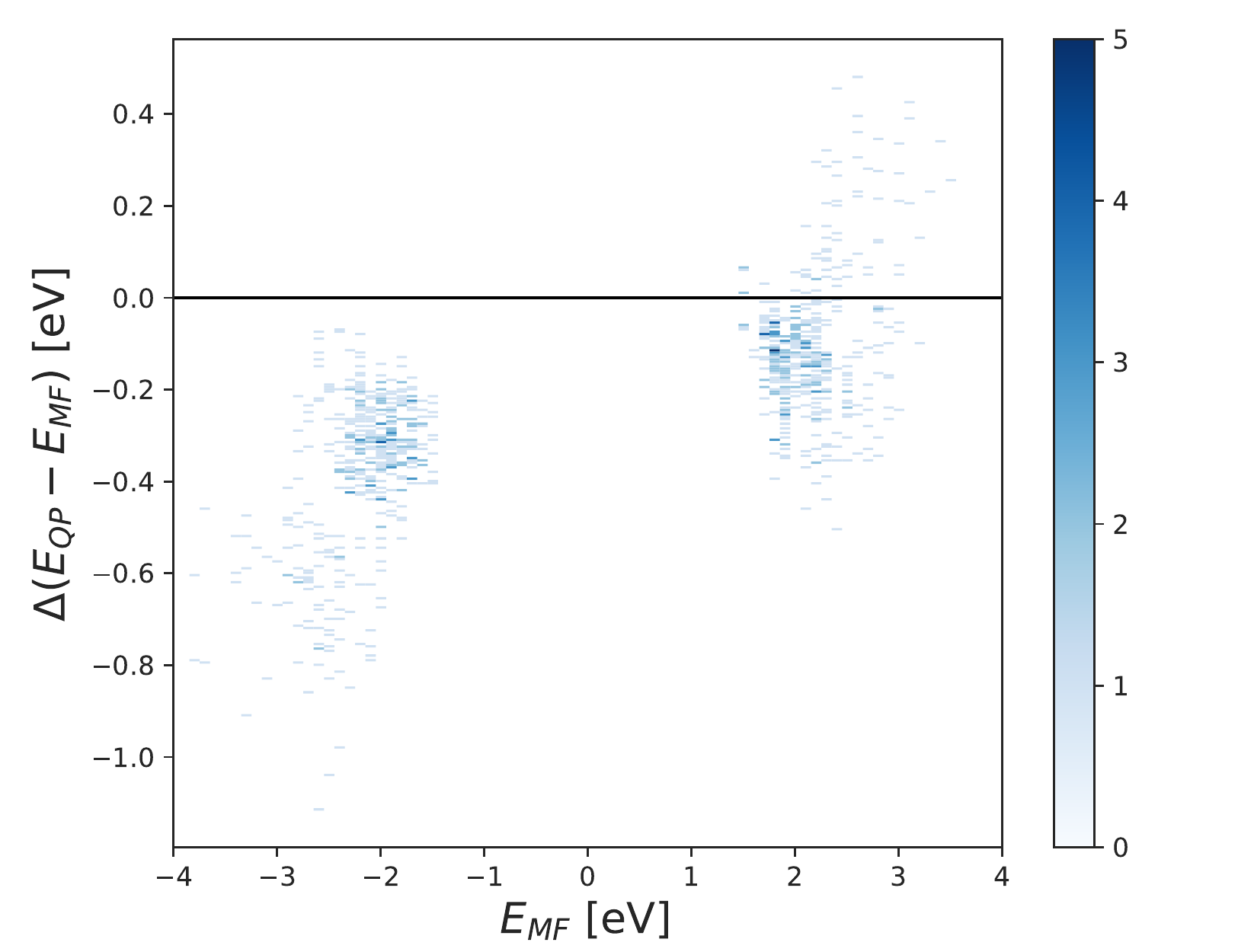}
          \end{subfigure}

    \caption{(a) Quasiparticle corrections from plane-wave G$_0$W$_0$ and G$_0$W$_0$ with the 1C-DDRF for the 10 highest valence orbitals and the 10 lowest conduction orbitals of hydrogenated silicon clusters.
    (b) Histogram of difference in quasiparticle corrections from plane-wave G$_0$W$_0$ and G$_0$W$_0$ calculations with the 1C-DDRF for the 10 highest valence orbitals and the 10 lowest conduction orbitals of hydrogenated silicon clusters. The mean-field energies are referenced to the middle of the mean-field HOMO-LUMO gap. %The energies of the smallest cluster were excluded. 0.29 eV excl. smallest cluster and 0.3 eV including smallest cluster.
    }    \label{qpcorrectionsdiff}
\end{figure}

Fig.~\ref{qpcorrectionsdiff} (a) shows the quasiparticle energy corrections of the ten lowest conduction orbitals and the ten highest valence orbitals from plane-wave G$_0$W$_0$ and G$_0$W$_0$ with the 1C-DDRF. The corrections obtained with the 1C-DDRF follow a similar trend as those obtained from the plane-wave calculation. For the unoccupied states, the quantitative agreement is better than for the occupied states, but the 1C-DDRF results for the unoccupied states are scattered over a larger energy range than the plane-wave results. To analyze the errors that arise from the use of the 1C-DDRF in more detail, Fig.~\ref{qpcorrectionsdiff} (b) shows a two-dimensional histogram of the difference in QP corrections between plane-wave G$_0$W$_0$ and G$_0$W$_0$ with the 1C-DDRF. For the occupied states the differences are mostly smaller than 0.4 eV, while they are somewhat smaller for the unoccupied states. The RMSE over all energy levels is 0.32 eV.

Now that we have established the accuracy of the method used to generate the training set, we use a dense neural network (NN) in conjunction with the NDM descriptor to generate the coefficients of the 1C-DDRF according to
\begin{equation}
    \chi^{(i)}_{nlmn'l'm'} = f(\rho^{(i, Si)}_{nlm}, \rho^{(i, H)}_{nlm}),
\end{equation}
where $f$ is the neural network function.
The hydrogen and silicon environment descriptors are concatenated into a single vector before being fed into the neural network. A separate network is trained for Si and H contributions to the DDRF. The exact architecture of the network as well as the practical computation of the atomic decomposition and the descriptors are described in the Methods section. To generate the training data for the neural network, we start from the set of relaxed hydrogenated Si clusters that were studied above. From each relaxed cluster, we generate six new configurations by randomly displacing the atoms with the magnitude of the displacements being drawn from a uniform distribution with a maximum of 0.1~\AA. For these clusters, we then calculate the 1C-DDRF.

\begin{figure}[h]
    \centering
    \includegraphics[scale=0.5]{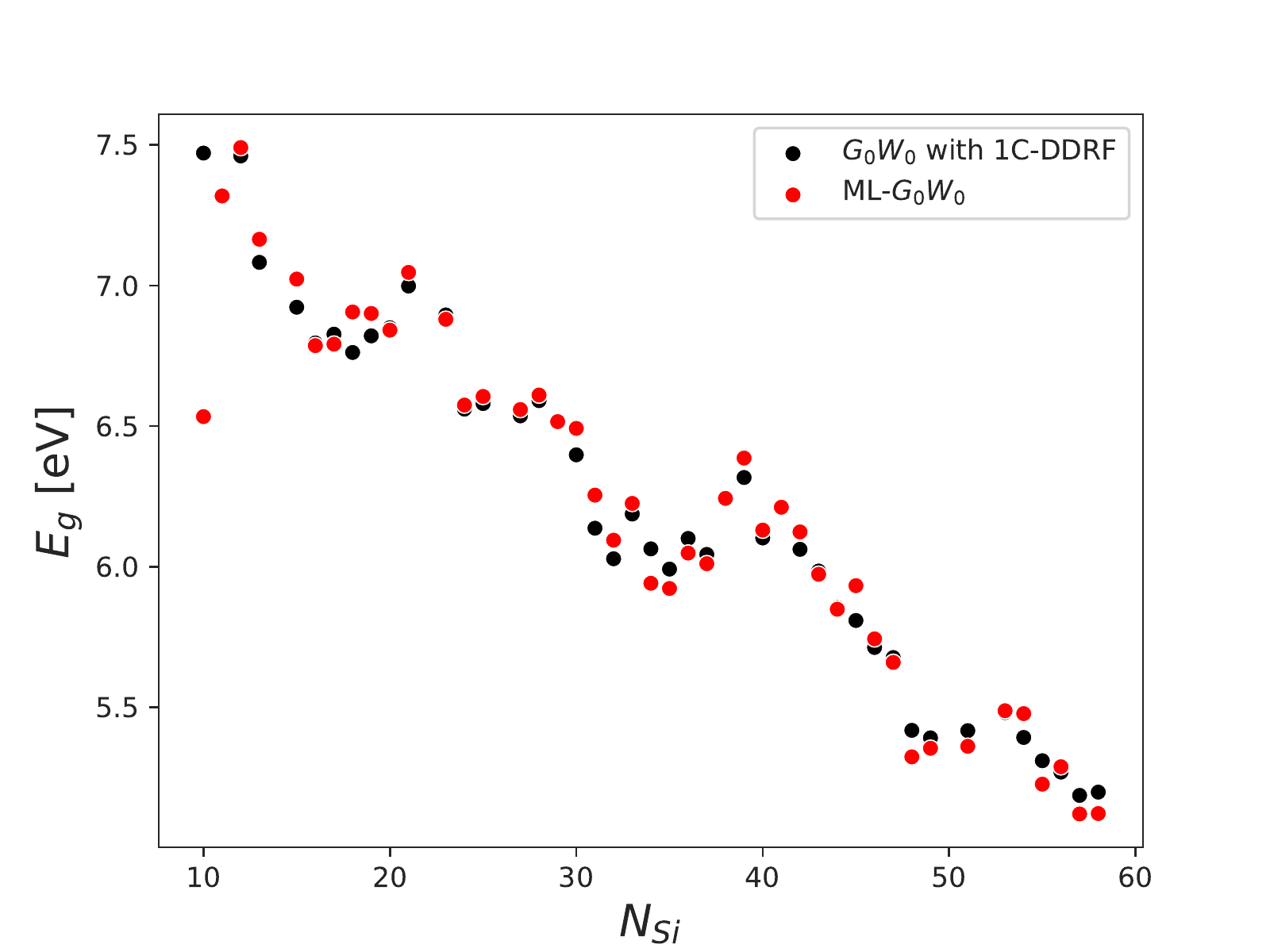}
    \caption{HOMO-LUMO gaps of hydrogenated silicon clusters from plane-wave G$_0$W$_0$ and G$_0$W$_0$ calculations using the 1C-DDRF and ML-G$_0$W$_0$. %0.15 eV including smallest cluster. 0.06 eV excluding smallest cluster.
    }
    \label{atomictrain}
\end{figure}

Once the neural network is trained on the 1C-DDRF of the randomly displaced clusters, we use it to calculate the 1C-DDRFs of the relaxed clusters and then determine quasiparticle energies via the ML-GW approach. Fig.~\ref{atomictrain} compares the HOMO-LUMO gaps from ML-GW and GW with explicitly calculated 1C-DDRFs. Except for the smallest cluster, the ML-GW method accurately reproduces the HOMO-LUMO gaps of the explicit GW calculations. The worse performance for the smallest cluster is a consequence of the training set which contains a large number of bigger clusters containing atomic environments that differ from those found in the smallest clusters. The overall RMSE of the ML-GW method relative to the explicit GW with the 1C-basis is only 0.15~eV, but reduces to 0.06 eV when the smallest cluster is excluded. 

\begin{figure}[h]
    \centering
    \includegraphics[scale=0.5]{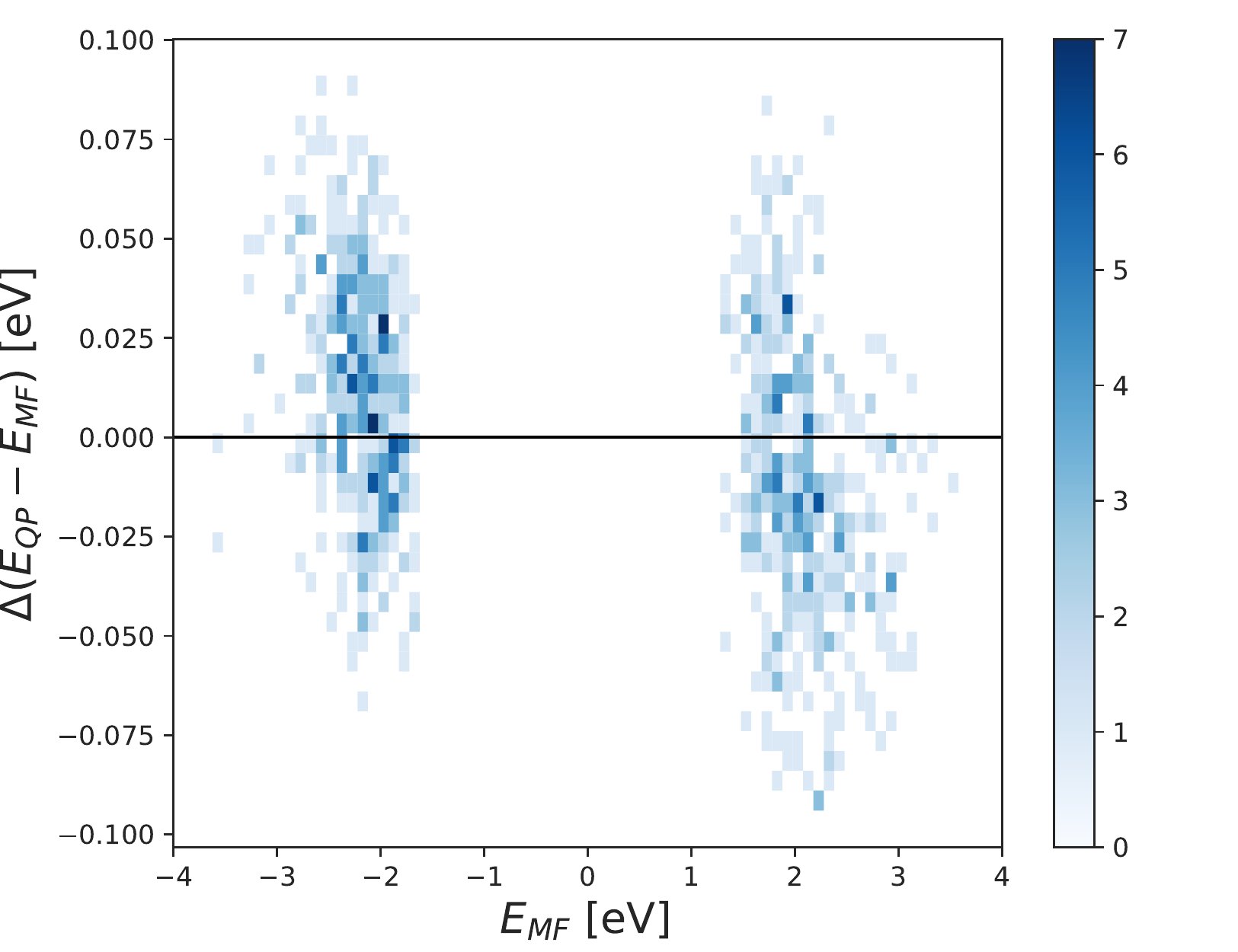}
    \caption{Histogram of difference in quasiparticle corrections from G$_0$W$_0$ using the 1C-DDRF and ML-G$_0$W$_0$ for the 10 highest valence orbitals and the 10 lowest conduction orbitals of hydrogenated silicon clusters. The mean-field energies are referenced to the middle of the mean-field HOMO-LUMO gap. The energies of the smallest cluster were excluded.}
    \label{qpcorrectionsdiff_baseatomic}
\end{figure}

Fig.~\ref{qpcorrectionsdiff_baseatomic} shows the difference in QP corrections between ML-GW and GW with the 1C-DDRF for the 10 highest valence states and 10 lowest conduction states. ML-GW produces QP shifts for both valence and conductions states within 0.1~eV from the explicit G$_0$W$_0$ with the 1C-DDRF. The majority of valence states exhibit a positive error, while for conduction states, the error is largely negative.

\begin{figure}[h]
    \centering    \includegraphics[scale=0.5]{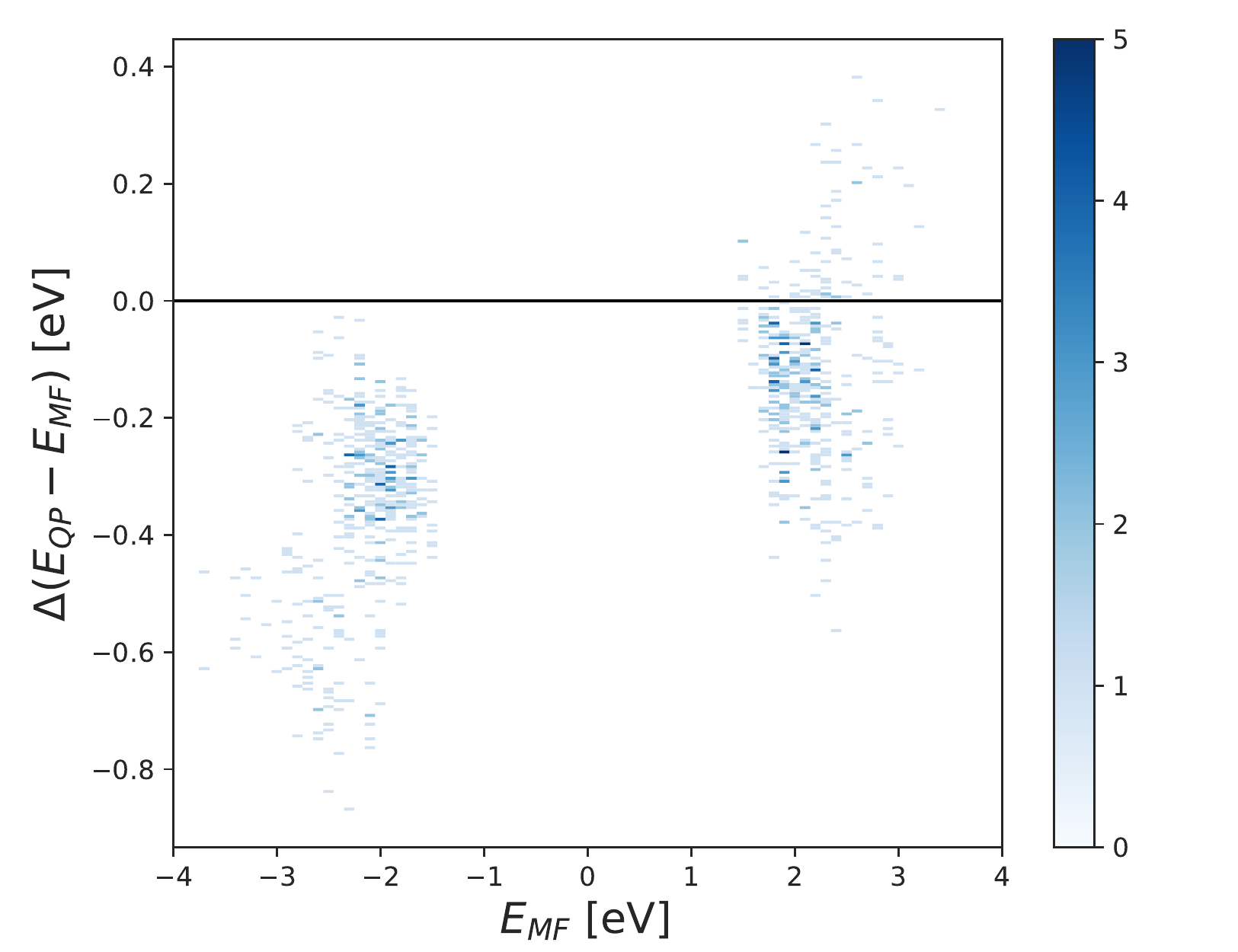}
    \caption{Histogram of difference in quasiparticle corrections from plane-wave G$_0$W$_0$ and ML-G$_0$W$_0$ DDRF for the 10 highest valence orbitals and the 10 lowest conduction orbitals of hydrogenated silicon clusters. The mean-field energies are referenced to the middle of the mean-field HOMO-LUMO gap. The energies of the smallest cluster were excluded. %0.33 eV excl. smallest cluster and 0.29 eV including smallest cluster.
    }    \label{qpcorrectionsdiff2}
\end{figure}

Fig.~\ref{qpcorrectionsdiff2} compares the ML-G$_0$W$_0$ QP corrections to plane-wave G$_0$W$_0$ results. As expected, the differences are very similar to those between plane-wave G$_0$W$_0$ and the explicit G$_0$W$_0$ with the 1C-basis. In particular, the RMSE is 0.35 eV for all clusters and reduces to 0.30 eV when the smallest cluster is excluded. This results demonstrates that the key obstacle to improving the ML-GW approach is the development of a better basis set.

\begin{figure*}[h!]
    \begin{subfigure}[t]{0.45\textwidth}
            \caption{}
    \includegraphics[width=0.97\textwidth]{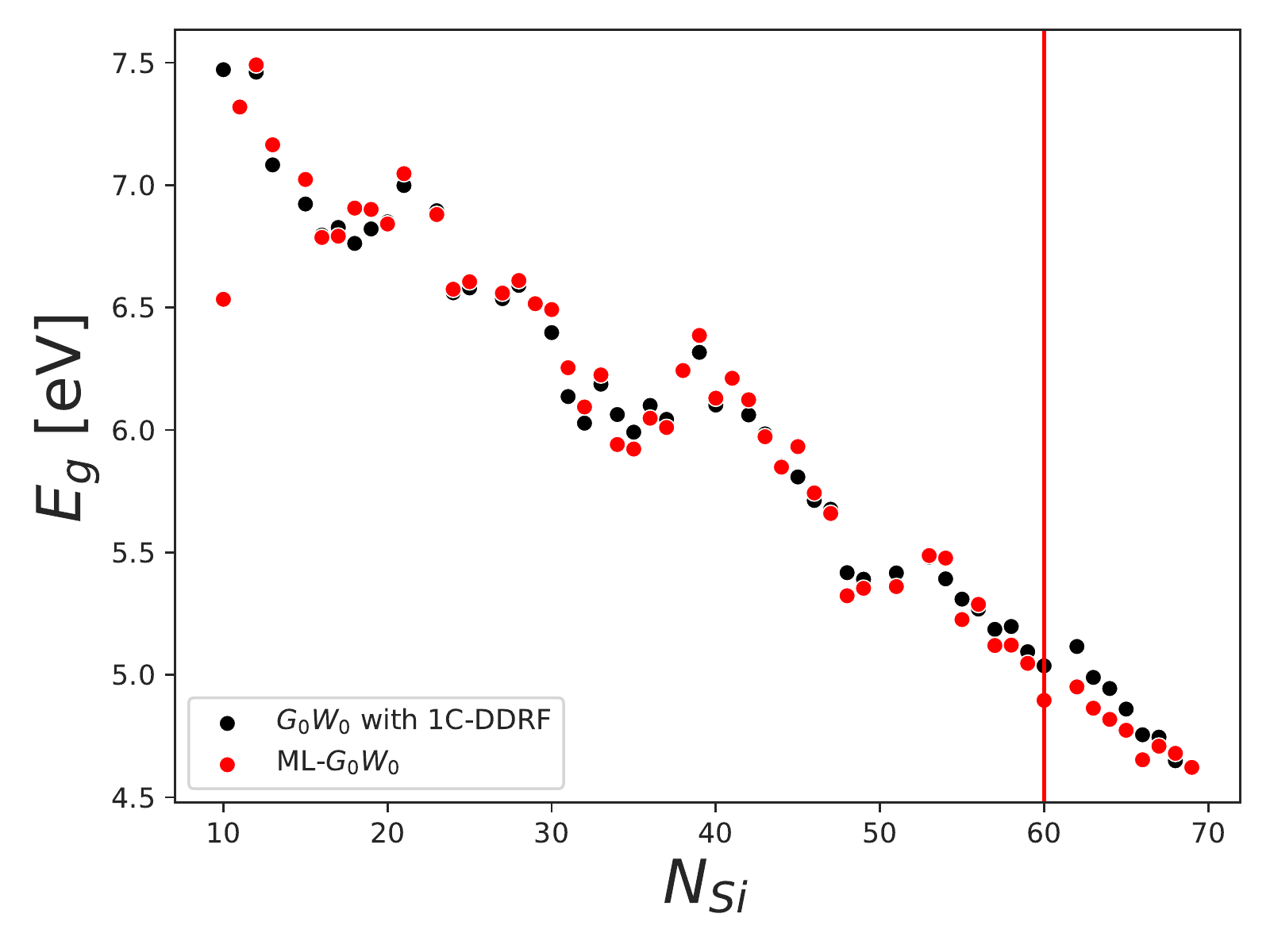}
    \end{subfigure}
         \begin{subfigure}[t]{0.47\textwidth}
                 \caption{}
    \includegraphics[width=\textwidth]{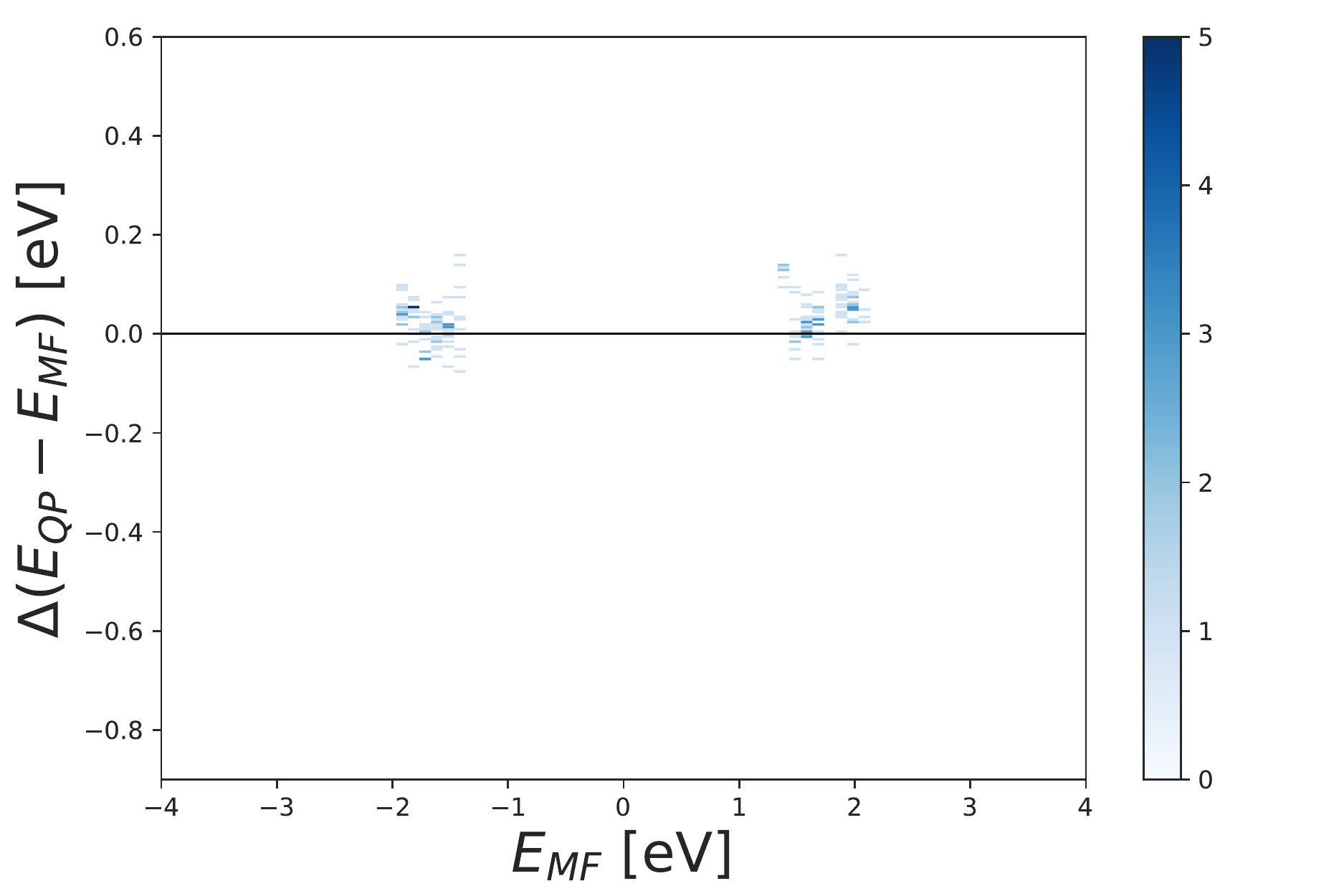}
    \end{subfigure}
    \begin{subfigure}[t]{0.45\textwidth}
        \caption{}
    \includegraphics[width=0.97\textwidth]{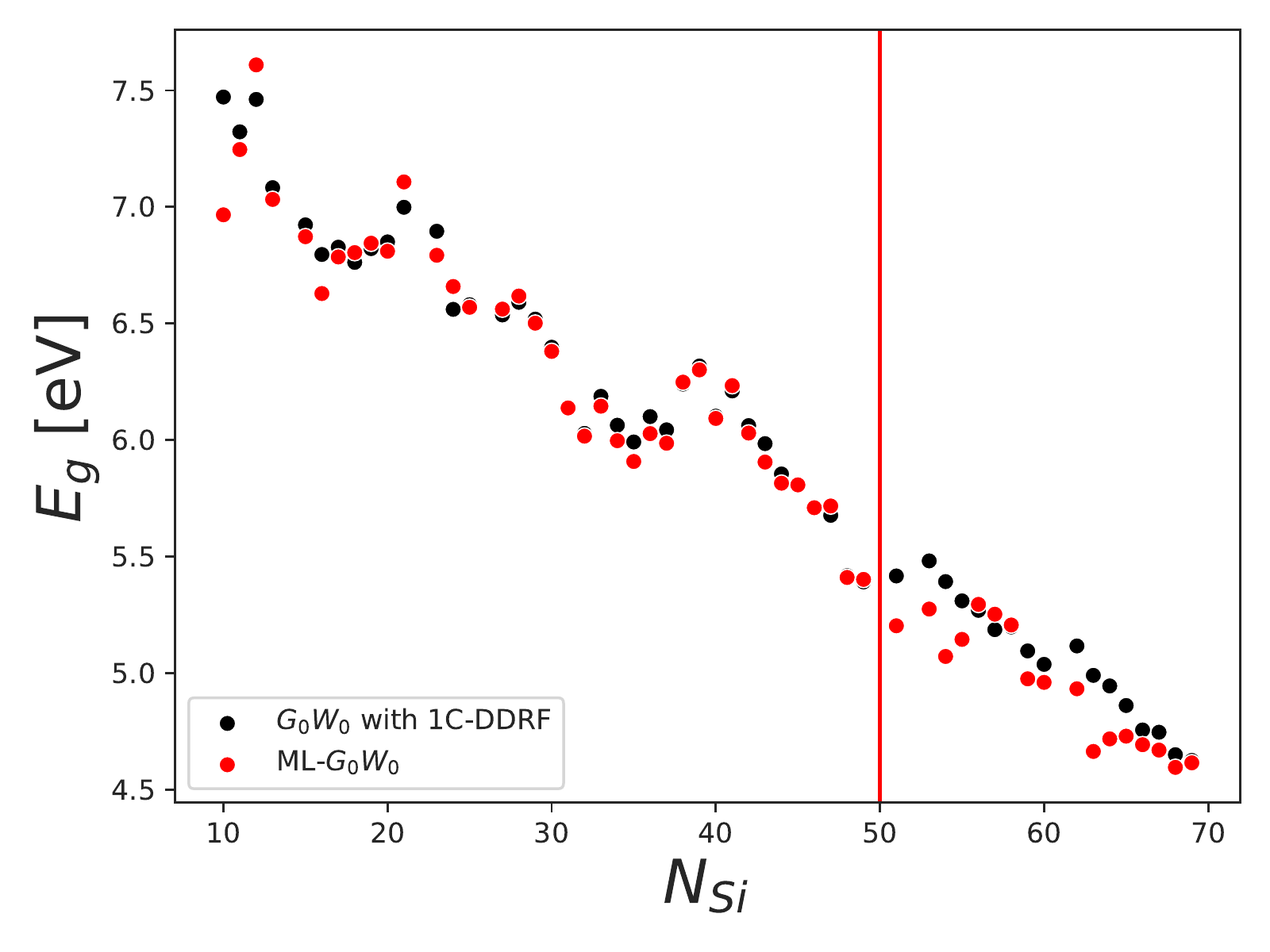}
    \end{subfigure}
        \begin{subfigure}[t]{0.47\textwidth}
          \caption{}
    \includegraphics[width=\textwidth]{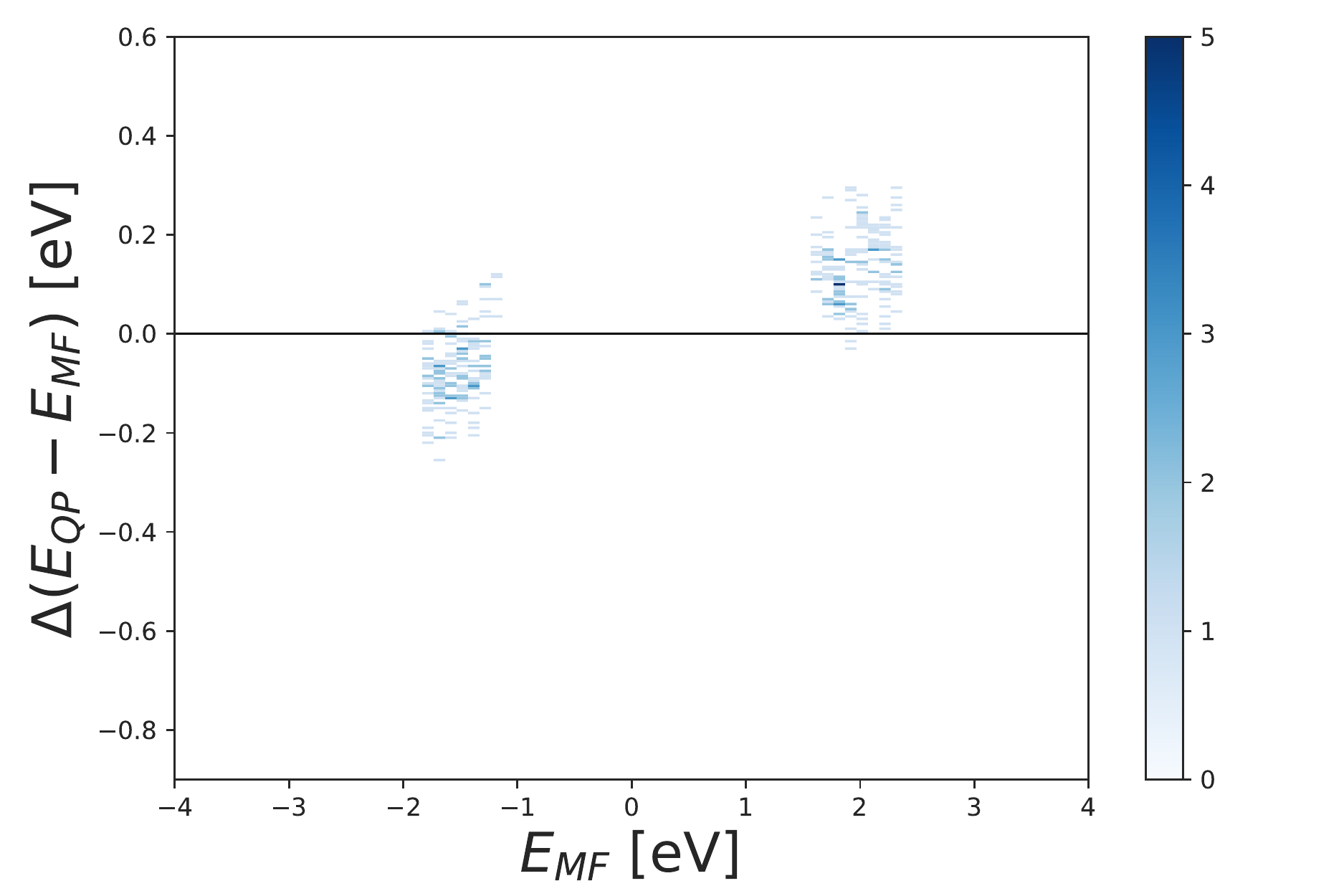}
    \end{subfigure}
    \begin{subfigure}[t]{0.45\textwidth}
     \caption{}
    \includegraphics[width=0.97\textwidth]{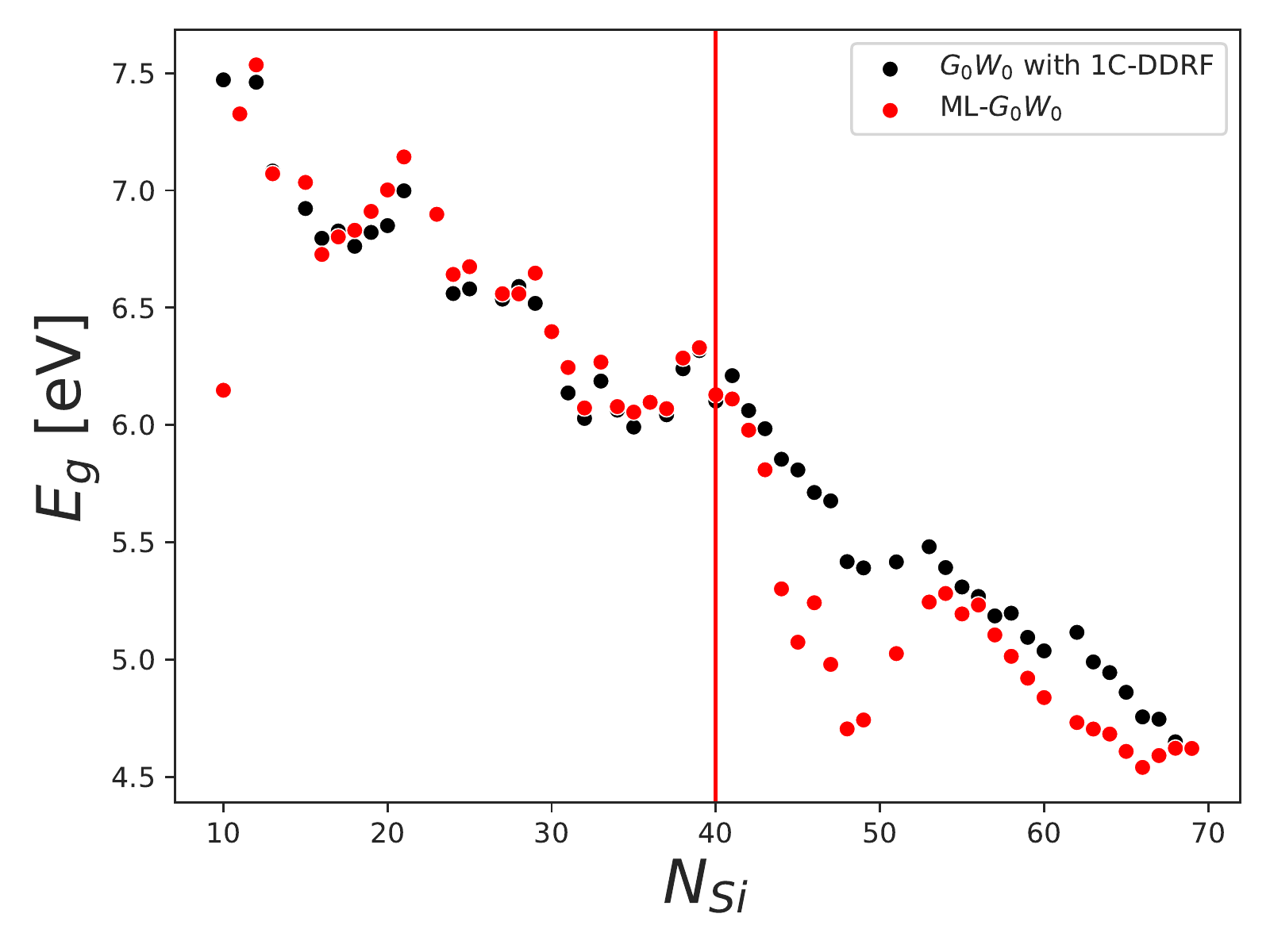}
    \end{subfigure}
    \begin{subfigure}[t]{0.47\textwidth}
      \caption{}
    \includegraphics[width=\textwidth]{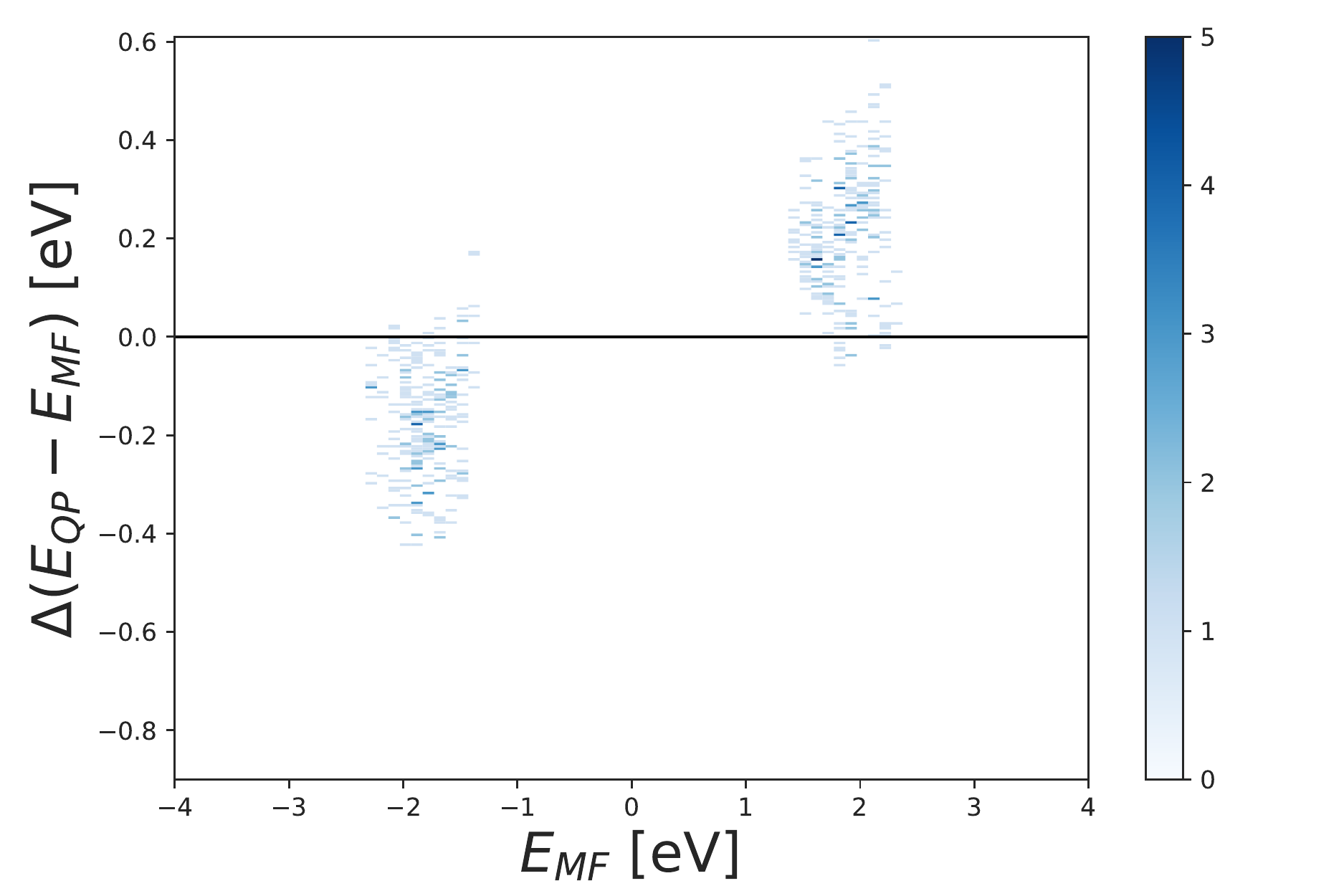}
    \end{subfigure}
     \caption{HOMO-LUMO gaps (left panels) and errors in quasiparticles shifts (right panels) from explicit G$_0$W$_0$ calculations with the 1C-DDRF and from ML-G$_0$W$_0$ trained on clusters containing up to $N_max=60$ Si atoms (upper panels), $N_max=50$ Si atoms (middle panels) and $N_max=40$ Si atoms (lower panels). The red vertical line indicates $N_max$. The panels on the right hand side only contain results for clusters with more Si atoms than $N_{max}$. The mean-field energies are referenced to the middle of the mean-field HOMO-LUMO gap.}
     \label{extr}
\end{figure*}

Finally, we test the ability of the ML-GW approach to predict the quasiparticle energies of clusters which are larger than those included in the training data. For this, we only include clusters with up to $N_{max}$ Si atoms in the training set with $N_{max}$ being 60, 50 and 40. Again, the training set only include clusters with randomly displaced atoms and the test set consists of the relaxed clusters. The predicted ML-GW for the whole set of relaxed clusters is shown in Fig.~\ref{extr}. From this graph, it is clear that the accuracy of the prediction for the largest clusters deteriorates as $N_{max}$ is reduced: while for $N_{max}=60$, the gaps and QP corrections for clusters with more than 60 Si atoms are still highly accurate, larger differences are observed for $N_{max}=50$. For $N_{max}=40$, errors as larger as 1 eV are obtained for the gaps of clusters with around 50 Si atoms. Fig.~\ref{extr}(f) shows that the large error in the gaps are a consequence of having a negative error in the QP shifts for occupied states and a positive error in the shift for unoccupied states. In other words: instead of a cancellation, we get an accumulation of errors when computing HOMO-LUMO gaps.

\section{Discussion}
We have developed a machine learning approach to predict the interacting density-density response function (DDRF) of materials. To achieve this, we introduce a decomposition of the DDRF into atomic contributions which form the output of a neural network. We also introduce the NDM descriptor which is a generalization of the widely used SOAP descriptor~\cite{Bartok2010}: instead of symmetrizing the descriptor using a Haar integral over a symmetry group~\cite{Langer2022}, we construct the tensor product of the expansion coefficients of the neighbourhood density which transforms under rotation in the same way as the atomic contributions to the DDRF. Thus, while not fully covariant, our approach is able to distinguish between different orientations of a chemical environment, which is a key requirement for predicting functions, such as the DDRF. 

The machine learning technique for DDRFs is then combined with the GW approach. The resulting approach is called the ML-GW approach. We apply this method to hydrogenated silicon clusters. The ML-GW approach reproduces HOMO-LUMO gaps and quasiparticle energies of GW calculations using the explicitly calculated 1C-DDRF, i.e. the DDRF in a pair basis where the basis functions of each pair are centered on the same atom, with an accuracy of about 0.1 eV. The accuracy of the results deteriorates when it is applied to clusters which are larger than those included in the training set. 

However, the error of ML-GW is significantly larger when compared to standard plane-wave GW results: HOMO-LUMO gaps are reproduced to within 0.5 eV, but the error reduces to 0.4 eV when the smallest cluster is excluded from the test set. These errors are comparable to those obtained by Rohlfing in his GW calculations for silane using a model dielectric function~\cite{Rohlfing}. 

These findings demonstrate that the main challenge towards improving the ML-GW method is the construction of better local basis sets for the DDRF. The basis used for the 2C-DDRF can be improved straightforwardly by using larger basis sets, such as aug-admm-2, admm-3 or aug-admm-3~\cite{Kumar2018}. However, it is more difficult to increase the basis used for the 1C-DDRF as this leads to linear dependencies which deteriorate the predictive accuracy of the neural network. This was also observed by Grisafi et. al. \cite{Grisafi2019} when predicting the expansion coefficients of the electronic density using the symmetry adapted SOAP kernel \cite{Grisafi2018}. In the future, we plan to explore the use of orthogonal radial basis sets, such as Laguerre polynomials instead of solid harmonic Gaussians. 

We expect that the ML-GW method can be applied to calculate quasiparticle energies in systems that have so far been out of reach for standard implementations. Examples include disordered materials, liquids, interfaces or nanoparticles. It could also be combined with on-the-fly machine learning methods~\cite{Kermode2015} to perform GW calculations on molecular-dynamics snapshots to determine finite-temperature quasiparticle energies.

\section{Methods}

\subsection{Data generation}
The atomic structures of the hydrogenated silicon clusters were obtained in the same way as described by Zauchner et. al. \cite{Zauchner_2021}: starting from the Si$_{123}$H$_{100}$ cluster of the silicon Quantum Dot data set~\cite{barnard2015silicon}, we remove the silicon atom furthest from the centre of the cluster, terminate the dangling bonds with hydrogen atoms and relax the resulting structure using DFT. The process is repeated until only 10 silicon atoms remain. In addition, we also include the clusters with less than 123 Si atoms from the silicon Quantum Dot data set. From this set of silicon clusters, only clusters with less than 60 silicon atoms were used in the training set for DDRF prediction. From each cluster with less than 60 silicon atoms, we created six additional clusters in which random displacements were added to the atomic positions. The magnitudes of the displacements were drawn from a normal distribution with mean of 0 \AA\ and standard deviation of 0.1 \AA. Finally, calculations were also carried out for 
clusters with between 60 and 70 silicon atoms. These clusters are not part of the training set, but are used to test the extrapolation capacity of the ML approach.

\subsection{DFT and GW calculations}
The DDRF and QP corrections were calculated using the BerkeleyGW software package~\cite{bgw1, bgw2}. Mean-field DFT calculations were performed using the Quantum Espresso code~\cite{Giannozzi_2017, Giannozzi_2009}. Norm-conserving pseudopotentials from the Quantum Espresso Pseudopotential Library were used. The parameters of the DFT calculations were the same as those used by Zauchner et. al. \cite{Zauchner_2021}: a plane-wave cut-off of 65 Ry, and a supercell with sufficient vacuum to avoid interactions between periodic images. 
For the calculation of the DDRF a total of 1000 Kohn-Sham states were used in the summation. Also, a plane-wave cut-off of 6 Ry and a truncated Coulomb interaction were used. The QP corrections were calculated using the generalized plasmon-pole approximation (GPP)~\cite{Hybertsen1986}, an explicit sum over 1000 Kohn-Sham states and also a static remainder correction~\cite{deslippe2013coulomb}. To calculate the HOMO and LUMO energies, the vacuum level was determined by averaging the electrostatic potential over the faces of the supercell.

\subsection{Projection onto intermediate basis}
We first use BerkeleyGW to calculate the inverse dielectric matrix $\epsilon^{-1}_{\mathbf{GG'}}$ in a plane-wave basis~\cite{bgw2}. From this, we determine the interacting DDRF via
\begin{equation}
    \chi_{\mathbf{GG'}} =  (\epsilon_{\mathbf{GG'}} - \delta_{\mathbf{GG'}})/v_{\mathbf{G}}
\end{equation}
with $v_\mathbf{G}$ being the Fourier transform of the truncated Coulomb interaction. 

Next, the DDRF in real space is obtained as
\begin{equation}
    \chi(\mathbf{r,r'}) = \frac{1}{V}\sum_{\mathbf{G,G'}} e^{i\mathbf{G\cdot r}} \chi_{\mathbf{GG'}} e^{-i\mathbf{G'\cdot r'}},
\end{equation}
where $V$ is the volume of the supercell.

Starting from a set of real atom-centred basis functions ${\phi}^{i}_{\alpha_i}(\mathbf{r})$, where $\alpha_i$ labels the basis function on atom $i$, we construct an orthogonal basis set $\tilde{\phi}^{i}_{\alpha_i}(\mathbf{r})$ 
\begin{equation}
    \tilde{\phi}^{i}_{\alpha_i}(\mathbf{r}) = \sum_{k} \sum_{\alpha_k} A_{ik}^{\alpha_i \alpha_k} {\phi}^{k}_{\alpha_k}(\mathbf{r}),
    \label{orthogonal}
\end{equation}
where $A_{ik}^{\alpha_i \alpha_k}$ is the matrix of eigenvectors of the overlap matrix. The coefficients of the DDRF when expanded in the orthogonalized basis are
\begin{multline}
\tilde{\chi}_{\alpha_i \alpha_j}^{ij} = \frac{1}{V}\sum_{\mathbf{G,G'}}  \chi_{\mathbf{G, G'}} \\ \times \int_{- \infty}^{{\infty}}  \tilde{\phi}^{i}_{\alpha_i}(\mathbf{r}) e^{i\mathbf{G\cdot r}} d\mathbf{r} \int_{- \infty}^{{\infty}}     e^{-i\mathbf{G'\cdot r'}} \tilde{\phi}^{j}_{\alpha_j}(\mathbf{r'}) d\mathbf{r'},
\end{multline}
where, due to the localised nature of the basis functions, we extended the integral from an integral over the supercell to an integral over all space. These integrals are proportional to the Fourier transforms of the basis functions (or their complex conjugates). 

We then transform back to the non-orthogonal localised basis set using Eq.~\eqref{orthogonal} to find
\begin{multline}
    \chi(\mathbf{r,r'}) = \sum_{\alpha_i \alpha_{j}} \sum_{ij} \tilde{\chi}_{\alpha_i \alpha_j}^{ij} \tilde{\phi}^{i}_{\alpha_i}(\mathbf{r})\tilde{\phi}^{j}_{\alpha_j}(\mathbf{r'}) = \\ \sum_{\alpha_k \alpha_{l}} \sum_{kl} \sum_{\alpha_i \alpha_{j}} \sum_{ij}  A_{ik}^{\alpha_i \alpha_k} A_{jl}^{\alpha_i \alpha_k}\tilde{\chi}_{\alpha_i \alpha_j}^{ij}{\phi}^{k}_{\alpha_k}(\mathbf{r}){\phi}^{l}_{\alpha_l}(\mathbf{r'})\\
    = \sum_{\alpha_k \alpha_{l}} \sum_{kl} 
    {\chi}_{\alpha_k \alpha_l}^{kl}{\phi}^{k}_{\alpha_k}(\mathbf{r}){\phi}^{l}_{\alpha_l}(\mathbf{r}'),
\end{multline}
where we defined
\begin{equation}
    {\chi}_{\alpha_k \alpha_l}^{kl} = \sum_{\alpha_i \alpha_{j}} \sum_{ij}
    \tilde{\chi}_{\alpha_i \alpha_j}^{ij}A_{ik}^{\alpha_i \alpha_k} A_{jl}^{\alpha_i \alpha_k}.
\end{equation}
The basis functions we employed are the real solid harmonic Gaussians as defined in LibInt2 \cite{Libint2}
\begin{equation}
    \phi_{lm}(r, \theta, \phi) = N_{l}(\beta) r^l e^{-\beta r^2} R_{lm}(\theta, \phi),
\end{equation}
where $\beta$ is a decay parameter, $N_l(\beta)$ is a normalization factor and $R_{lm}$ are the real spherical harmonics given by \cite{Schlegel1995TransformationBC}
\begin{align}
    & R_{lm}(\theta, \phi) = \\ &\begin{cases}
     \frac{i}{\sqrt{2}} \left(Y_{l-\vert m \vert }(\theta,\phi)-(-1)^m Y_{l\vert m \vert }(\theta,\phi) \right) \text{ if } m<0 \\
      Y_{lm}(\theta, \phi) \text{ if } m=0\\
      \frac{1}{\sqrt{2}} \left(Y_{l-\vert m \vert }(\theta,\phi)+(-1)^m Y_{l\vert m \vert }(\theta,\phi) \right) \text{ if } m>0,
    \end{cases}
    \label{real}
\end{align}
where $Y_{l m  }(\theta,\phi)$ are the complex spherical harmonics with the Condon-Shortley phase convention.
Kuang and Lin showed that the Fourier transform of the complex solid harmonic Gaussians is again a solid harmonic Gaussian \cite{Kuang_1997}
\begin{multline}
    \frac{1}{(2\pi)^{3/2}} \int d\mathbf{r} e^{-i\mathbf{G\cdot r}} N_{l}(\beta) r^l e^{-\beta r^2} Y_{lm}(\hat{\mathbf{r}})\\ =(-i)^l \tilde{N}_{l}(\beta) G^l e^{- G^2/(4\beta)} Y_{lm}(\hat{\mathbf{G}}),
    \label{FT}
\end{multline}
with $\tilde{N}_{l}(\beta) = N_{l}(\beta)/(2\beta)^{3/2}$.
The Fourier transform of the real solid harmonic Gaussians can then be easily computed using Eq.~\eqref{real}. 

The basis set used in this work is a modified version of the admm-2 basis set \cite{Kumar2018} (see Appendix for details), in which the s-orbitals were removed and contracted Gaussians were uncontracted into individual basis functions. Removing the s-orbitals ensures that $\int d\mathbf{r} \chi(\mathbf{r},\mathbf{r}')=0$ since only the Fourier transform of s-orbitals has a $\mathbf{G}=0$ contribution.

%The removal of s-type orbitals is that only s-type orbitals have a non-zero contribution at $\mathbf{G}=0$, as can be seen from Equation \ref{FT}. Removing s-orbitals ensures that the total DDRF integrates to zero. This property is a formal requirement, since the DDRF can be used to compute the density response to a perturbing potential \cite{Onida2002}
%\begin{equation}
%    \Delta \rho(\mathbf{r}) = \int \chi(\mathbf{r,r'}) V_{pert}(\mathbf{r'}) d\mathbf{r'},
%    \label{response}
%\end{equation}
%which has to integrate to zero for any perturbing potential in order to be a true density response. 
%This means that in Fourier-space $\chi_{\mathbf{0 G'}}=0 $ $ \forall \mathbf{G'}$. Further, using the symmetry property of the DDRF 
%\begin{equation}
%\chi(\mathbf{r,r'})= \chi(\mathbf{r',r}),
%    \label{symmetry}
%\end{equation}
%combined with the static response being real, means that $\chi_{\mathbf{G G'}} = \chi_{\mathbf{G' G}}^*$. Thus, ensuring that both the first row $\chi_{\mathbf{0 G'}}$ and the first column $\chi_{\mathbf{G 0}}$ of $\chi$ are zero enforces that the response to any perturbing potential integrates to zero. 

%We note that including s-orbitals in the intermediate basis, and afterwards zeroing out the first row and first column of $\chi_{\mathbf{G G'}}$, yields slightly more accurate QP energies in the intermediate basis. However we found that the projection onto the atomic basis described in the next section becomes less stable when including s-orbitals in the intermediate basis.

\subsection{Projection onto atomic basis}

The fully atom-centred basis set also consists of solid harmonic Gaussians. The basis set was constructed following the same procedure as in the DScribe library \cite{dscribe}, where individual basis functions are given by 
\begin{equation}
      \psi_{nlm}(r, \theta, \phi) = N_{l}(\beta_{nl}) r^l e^{-\beta_{nl} r^2} R_{lm}(\theta, \phi),
\end{equation}
where the basis set is truncated at a maximum angular momentum $l_{max}$ and a maximum principal quantum number $n_{max}$. For silicon atoms we use $l_{max} = n_{max}= 4$. For hydrogen atoms we use $l_{max} = n_{max}= 3$.

The exponents $\beta_{nl}$ are constructed such that the corresponding basis functions decay to zero at a cutoff radius $R_n$, i.e. $
    \beta_{nl} = - \ln(\frac{T}{R_n^l})/R_n^2
$
with $T=10^{-3}$ \AA$^l$ being a threshold parameter. The cutoff radius $R_n=R_i + (R_o-R_i)/n$ lies between an inner radius $R_i$ and an outer radius $R_o$. For hydrogen atoms, we used $R_i=0.1$ \AA\ and $R_o= 3.0$ \AA\ and for silicon atoms, we used $R_i= 1.0$ \AA\ and $R_o=8.0$ \AA. Both $R_i$ and $R_o$ were optimized to minimize linear dependencies in the basis set as such dependencies significantly deteriorate the accuracy of the neural network predictions. A similar observation was made by Grisafi et. al. \cite{Grisafi2019} when learning electron densities, although a different approach was taken to remedy this issue in their work. 

In order to compute the coefficients of the atomic contributions to the DDRF in the fully atom-centered basis the same procedure as in the intermediate basis was used: the basis was first orthogonalized by computing the eigenvectors of the overlap matrix. Then the atomic DDRFs in the intermediate basis were projected onto the orthogonalized fully-atom centred basis with overlaps between the different basis functions being computed using LibInt-2~\cite{Libint2}. Then the atomic DDRFs were transformed back to the non-orthogonal basis produced the desired coefficients $\chi^{(i)}_{nlmn'l'm'}$. 

%One drawback of the above partitioning, followed by a projection onto atom centred basis functions, is that the number of coefficients included in the summation has to be truncated. This leads to a basis set error that will be introduced in both the intermediate basis, as well as the atom centred basis. However, doing so allows the evaluation of all involved integrals in closed form, e.g. through the Obara-Saika scheme \cite{Obara} or the method described by Kuang et. al. \cite{Kuang_1997}. 
%\begin{multline}
% \chi_i(\mathbf{r,r'}) 
% = \frac{1}{2}  \sum_{\alpha_n} \sum_{w, \alpha_w} \\
% \left\{  \chi^{nw}_{\alpha_n\alpha_w}   \phi_{\alpha_n}^{n}(\mathbf{r})  
%        \phi_{\alpha_v}^{v}(\mathbf{r'})
%        +  \chi^{wn}_{\alpha_w\alpha_n}  \phi_{\alpha_w}^{w} (\mathbf{r})\phi_{\alpha_n}^{n} (\mathbf{r'}) \right\}.
 %       \label{crossiteb}
%\end{multline}

\subsection{Descriptors}
The basis set for neighbourhood densities was generated using the same procedure as for the fully atom-centered basis for the DDRF. However, s-orbitals were not removed and the basis functions of the admm-2 basis set were not included. We used $R_i=1.0$ \AA\ for both hydrogen and silicon atoms and $R_o=4.0$ \AA\ for hydrogen atoms and $R_o=9.0$ \AA\ for silicon atoms. The exponents of the Gaussians in Eq.~\eqref{desc} were set such that the standard deviation of the Gaussians is 0.5 \AA. LibInt-2 \cite{Libint2} was again used to compute the required integrals for the projection. 

\subsection{Neural network}
A dense neural network with four hidden layers with 2000, 1500, 1000 and 2000 nodes respectively was constructed for both silicon and hydrogen atoms. Each layer uses a Leaky-ReLu activation function with a leak parameter of 0.1. The output layer was further symmetrized by adding its transpose. The loss used was the mean-squared error between the predicted and true expansion coefficients $\chi^{(i)}_{nlmn'l'm'}$. The neural network was trained on the perturbed clusters for 20,000 epochs. We found that adding dropout to the layers does not significantly improve the quasiparticle energies resulting from the predictions which is likely due to the similarity between the atomic environments in the training and test set.
\section{Data availability statement}
The data that support the findings of this study are available upon reasonable request from the authors.

\section{Acknowledgements}
This work was supported through a studentship in the Centre for Doctoral Training on Theory and Simulation of Materials at Imperial College London funded by the EPSRC (EP/L015579/1). We acknowledge the Thomas Young Centre under grant number TYC-101. This work used the ARCHER2 UK National Supercomputing Service via J.L.’s membership of the HEC Materials Chemistry Consortium of UK, which is funded
by EPSRC (EP/L000202).

\section*{References} 
\bibliography{sample}
\bibliographystyle{naturemag}

\section{Appendix A}
\subsection{Transformation of atomic density-density response function and NDM under rotation}

We study the transformation properties of the atomic contributions to the DDRF under the action of a representation of the rotation group $D(\hat{R})$ \cite{jeevanjee2011introduction, rose2013elementary}: 
\begin{equation}
    D({\hat{R}})\otimes D^{\dagger}({\hat{R}})\chi_i(\mathbf{r,r'}) = \chi_i(\hat{R}^{-1}\mathbf{r},\hat{R}^{-1}\mathbf{r'}).
    \label{transform}
\end{equation}
The transformation is defined by the tensor product representation $D({\hat{R}})\otimes D^{\dagger}({\hat{R}})$ of two SO(3) representations. Using 
the expansion of $\chi_i$ in the $\psi$-basis, we find
\begin{multline}
    \chi_i(\hat{R}^{-1}\mathbf{r},\hat{R}^{-1}\mathbf{r'}) = \sum_{nlm}\sum_{n'l'm'} \chi^{(i)}_{nlmn'l'm'} \\R_n(r)R_{n'}^{*}(r') Y_{lm}(\hat{R}^{-1}\mathbf{\hat{r}})Y_{l'm'}^{*}(\hat{R}^{-1}\mathbf{\hat{r}'})\\\\
    = \sum_{nlm}\sum_{n'l'm'} \chi^{(i)}_{nlmn'l'm'} R_n(r)R_{n'}^{*}(r')  \\ \sum_{m_1, m_2} D^{l*}_{mm_1}(\hat{R}^{-1})D^{l'}_{m'm_2}(\hat{R}^{-1}) Y_{lm_1}(\mathbf{\hat{r}})Y_{l'm_2}^{*}(\mathbf{\hat{r}'}).
\end{multline}

Exchanging the summations of $m, m'$ and $m_1, m_2$ and using the unitarity property of the Wigner-D matrices $D^{l}(\hat{R})$~\cite{rose2013elementary} yields
%\begin{equation}
%D^{l}(\hat{R}^{-1}) = D^{l\dagger}(\hat{R}),
%    \label{unitary}
%\end{equation}
%we arrive at
\begin{multline}
    \chi_i(\hat{R}^{-1}\mathbf{r},\hat{R}^{-1}\mathbf{r'})=\\
\sum_{nlm_1}\sum_{n'l'm_2} R_n(r)R_{n'}^{*}(r')  Y_{lm_1}(\mathbf{\hat{r}})Y_{l'm_2}^{*}(\mathbf{\hat{r}'})\\ \sum_{m, m'} D^{l}_{m_1m}(\hat{R})D^{l'*}_{m_2m'}(\hat{R}) \chi^{(i)}_{nlmn'l'm'}.
\end{multline}
Thus, the coefficients of the atomic DDRF transform under rotation as (with the tilde denoting the rotated coefficients)
\begin{equation}
\tilde{\chi}^{(i)}_{nlm_1n'l'm_2} =      \sum_{m, m'} D^{l}_{m_1m}(\hat{R})D^{l'*}_{m_2m'}(\hat{R})\chi^{(i)}_{nlmn'l'm'}
\end{equation}
or in matrix notation
\begin{equation} \tilde{\chi}^{(i)}_{nn'll'} = \mathbf{D}^{l}(\hat{R})\mathbf{\chi}^{(i)}_{nn'll'}\mathbf{D}^{l'\dagger}(\hat{R}).
\end{equation}

By going through the same steps, it can be shown that the expansion coefficients of neighbourhood density matrix obey
\begin{equation}
\tilde{\rho}^{(i, \eta)}_{nlm_1n'l'm_2} =      \sum_{m, m'} D^{l}_{m_1m}(\hat{R})D^{l'*}_{m_2m'}(\hat{R}) \rho^{(i, \eta)}_{nlmn'l'm'}
\end{equation}
under rotation.
\section{Appendix B}
\subsection{Exponents of modified admm-2 basis}
The exponents of the modified admm-2 basis sets for hydrogen and silicon atoms are shown in Tables \ref{hydrogenbasis} and \ref{siliconbasis}. The basis functions were normalized to 1.
\begin{table}[h!]
\caption{Hydrogen basis}
\begin{tabular}{|l|l|}
\hline
$l$ & $\beta$ [$1/a_0^2$]\\ \hline
1   & 1.0      \\ 
1    & 0.457639 \\ \hline 
\end{tabular}
\label{hydrogenbasis}
\end{table}
\begin{table}[h!]
\caption{Silicon basis}
\begin{tabular}{|l|l|}
\hline
\multicolumn{1}{|l|}{$l$} & \multicolumn{1}{l|}{$\beta$ [$1/a_0^2$]} \\ \hline
1                         & 13.8028             \\
1                          & 59.9261                       \\
1                         & 4.34446                       \\
1                         & 0.267360                      \\
1                         & 0.0765250                     \\
2                         & 0.45                  \\       \hline
\end{tabular}
\label{siliconbasis}
\end{table}
\end{document}